\documentclass[review]{elsarticle}
\usepackage{amsmath}
\usepackage{amssymb}
\usepackage{booktabs}
\usepackage{pifont} 
\usepackage{lineno,hyperref}
\usepackage{longtable}
\usepackage{float}
\usepackage{array}
\usepackage{booktabs}
\usepackage{hyperref}
\usepackage{placeins}

\journal{Indian Institute of Technology - Kanpur}

\begin{document}

\begin{frontmatter}

\title{Heads Up eXperience (HUX): Always-On AI Companion for Human Computer Environment Interaction}

\author{Sukanth K \fnref{Corresponding author}}
\address{Vellore Institute of Technology - Chennai, Tamil Nadu, India}
\fntext[myfootnote]{Corresponding author}
\ead{sukanth.k2021@vitstudent.ac.in}

\author{Sudhiksha Kandavel Rajan}
\address{Shiv Nadar University - Chennai, Tamil Nadu, India}
\ead{sudhiksha21110126@snuchennai.edu.in}

\author{Rajashekhar V S}
\address{Indian Institute of Technology - Kanpur, Uttar Pradesh, India}
\ead{raja23@iitk.ac.in}

\author{Gowdham Prabhakar}
\address{Indian Institute of Technology - Kanpur, Uttar Pradesh, India}
\ead{gowdhampg@iitk.ac.in}

\begin{abstract}
While current personal smart devices excel in digital domains, they fall short in assisting users during human environment interaction. This paper proposes Heads Up eXperience (HUX), an AI system designed to bridge this gap, serving as a constant companion across the extended reality (XR) environments. By tracking the user's eye gaze, analyzing the surrounding environment, and interpreting verbal contexts, the system captures and enhances multi-modal data, providing holistic context interpretation and memory storage in real-time task specific situations. This comprehensive approach enables more natural, empathetic and intelligent interactions between the user and HUX AI, paving the path for human computer environment interaction. Intended for deployment in smart glasses and extended reality headsets, HUX AI aims to become a personal and useful AI companion for daily life. By integrating digital assistance with enhanced physical world interactions, this technology has the potential to revolutionize human-AI collaboration in both personal and professional spheres paving the way for the future of personal smart devices.
\end{abstract}


\end{frontmatter}

\begin{figure}[H]
    \centering
    \includegraphics[scale=0.12]{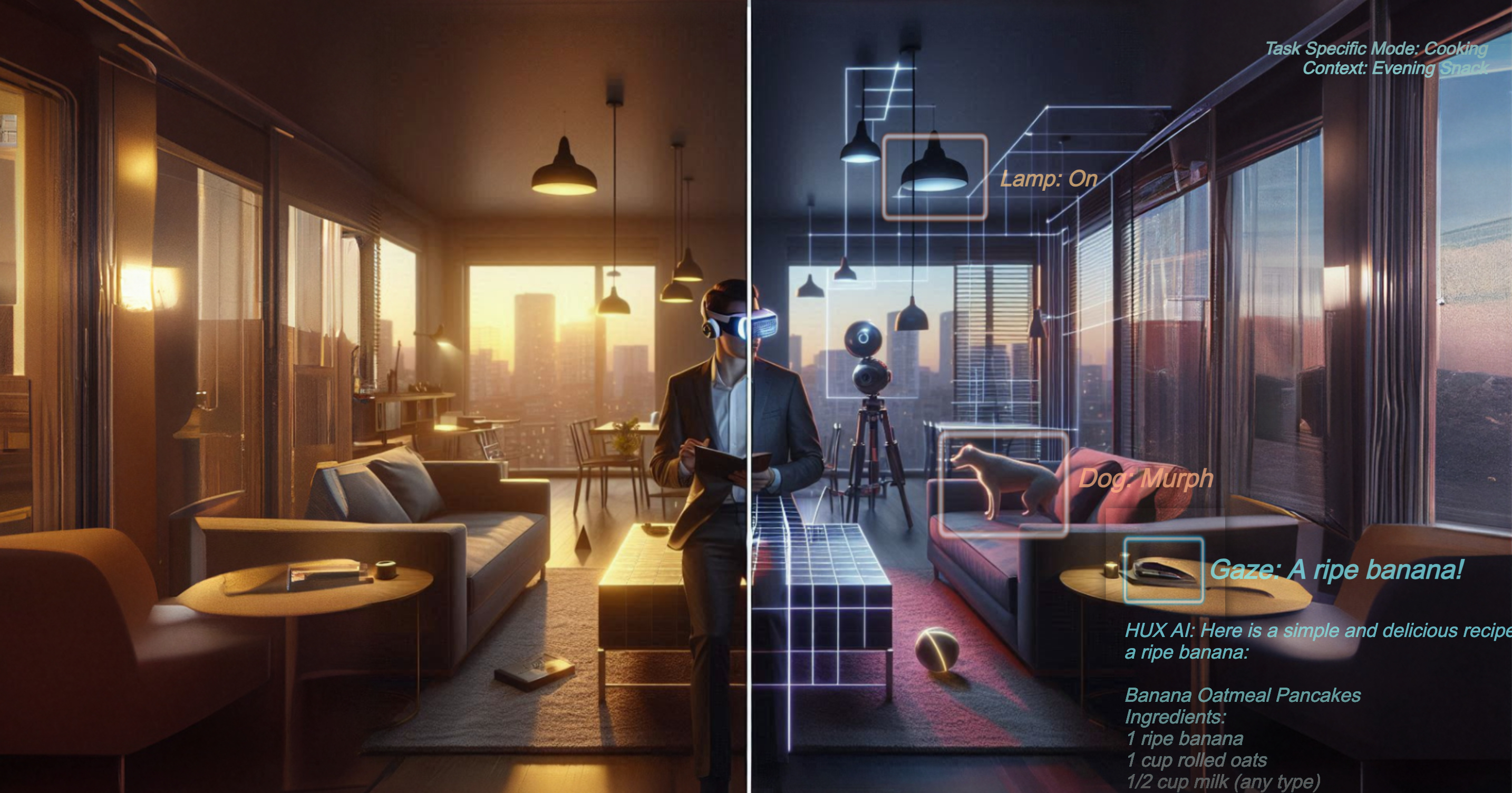}
    \caption{Concept of HUX AI. You can watch the video on YouTube by following this link (https://youtu.be/rM3la0N6vKM)}
    \label{fig:teaser}
\end{figure}

\section*{Nomenclature}
\begin{itemize}
\item AI - Artificial Intelligence
\item AR - Augmented Reality
\item XR - Extended Reality
\item PSD - Personal Smart Devices
\item HCI - Human Computer Interaction
\item HCEI - Human Computer Environment Interaction
\item VLM - Vision Language Model
\item LLM - Large Language Model
\item LAM - Large Action Model
\item GUI - Graphical User Interface
\item HUX - Heads-Up eXperience
\item OOI - Object of interest
\item EOI - Event of interest
\item ROI - Region of interest
\item PCB - Printed Circuit Board
\end{itemize}

\section*{Definitions}
\begin{itemize}

\item {\texttt{HUX AI}}:  The HUX AI is designed to realize the principles of HCEI.

\item {\texttt{HUX Device}}: A device capable of deploying the HUX AI. Having all sets of input and output modalities necessary for gathering multi-modal user inputs (speech, gaze), environment (scene) inputs, and multi-modal outputs (display and speech) such as a custom setup, smart glass, and extended reality headsets.

\item {\texttt{Task of interest}}: A task of interest is a domain-specific task the user wants to accomplish when using or wearing the HUX device.

\item {\texttt{Objects of interest (OOI)}}: In a task-specific scenario, objects of interest are the objects that fit within the context of the sequences of the task. Tracking its existence, position, behavior, and more are essential because they directly or indirectly affect the outcome of the task.

\item {\texttt{Events of interest (EOI)}}: In a task-specific scenario, events of interest are significant occurrences characterized by changes in the behavior of one or more objects within a specific scenario. These events are crucial for analysis as they help reconstruct and understand the sequence of temporal changes and their impacts.

\item {\texttt{Real-Time Video Analyzer}}: A real-time video analyzer is a frame-by-frame analyzer that employs computer vision models to detect objects and events of interest within a video feed. It extracts frames that capture events of interest related to the objects of interest in a specific scenario. In addition, it visually labels the real-time video feed to highlight the locations of objects of interest within the environment (if needed).

\item {\texttt{Multi-modal Context Processing}}: It has a prompt architecture that inserts multi-modal data in its text representations. This prompt is fed into an LLM instance, which outputs informative, context-aware answers, mode (data) aware, and decisive. 

\item {\texttt{Region of Interest (ROI)}}: It is the region in the scene where the user's selective attention is focused. It can be extracted by objects of interest specified verbally or by eye gaze, which is done in this work. 

\end{itemize}

\section{Introduction}
The usage of PSD plays an essential role in people's lives, and this trend has been observed in the past few decades. One of the primary focus is improving HCI. The next generation of personal smart devices (PSD)  has to be designed to improve interaction, helping humans in digital domains and the real world or connecting both. The present devices provide human computer interaction (HCI), but the future PSD should be capable of understanding human environment interaction and thus enabling human computer environment interaction (HCEI). This can be made possible by improving the input and output modalities in the current PSD.

The technological advancements enabled by transformer-based language \cite{brown2020language} and vision models seem to mimic human-like contextual understanding, as shown by their demonstrated capabilities \cite{street2024llms}. By harnessing these models, as mentioned above, and the multiple input modalities, we create a user-centric AI system whose goal is to enhance the HCEI.

The transformer-based models have already been deployed in commercial devices like Rabbit R1 \cite{rabbit2024}, Humane AI Pin \cite{humane2024}, and existing smartphones. However, choosing the right body to deploy the mind should be an essential design choice. The extended reality headset or smart glasses are the right choice because of their closer proximity to human perception and human communicative sense organs, which enhances the system's ability to align its perception with how users perceive the world. This alignment reduces friction in HCI because the AI system has access to user perception data. Consequently, it minimizes the need for human articulation, paving the way for intuitive communication.

The paper delves into the design of sub-processes serving unique purposes and integrates all of them to assemble a prototype. The goal of this prototype is to build a HCEI system. This work will aid in building the PSD with integrated AI—the current HCEI leverages multi-modal and real-time interaction. However, our system has taken a step forward by questioning the necessity of the current graphical user interface (GUI)-centric approach in PSD. The rest of the paper is organized as follows: A comparative analysis with existing literature is done in Section \ref{sec_past_works}. The problem statement is presented in Section \ref{sec:problem_state}. The heads-up experience (HUX) AI architecture is described in Section \ref{sec:hux_ai}. The applications of HUX AI are explained in Section \ref{sec:applications}. The experiments are presented in Section \ref{sec:experiments}. The discussion about the HUX AI is done in Section \ref{sec:discuss}. Section \ref{sec:lim_fut} presents the limitations and future works. Finally, the concluding remarks are presented in Section \ref{sec:conclu}. 

\section{Comparative Analysis with Existing Literature}
\label{sec_past_works}
In this section, we discuss the works from the literature. It is in selective attention theory, video processing, merits of visual enhancement in extended reality, embodied AI, and multi-modal HCI.

\subsection{Selective Attention Theory}
Selective attention is a critical cognitive process that allows individuals to focus on specific stimuli while ignoring irrelevant information. This concept is foundational in understanding human information processing and has been extensively studied through various models. Broadbent Filter Model (1958) \cite{broadbent2013perception}, for example, introduced the idea of an early-selection filter that processes inputs based on their physical characteristics. Treisman Attenuation Model (1964) \cite{treisman1964monitoring} refined this by suggesting that unattended messages are attenuated rather than completely filtered out, allowing some level of further processing. Deutsch and Deutsch Late Selection Theory (1963) \cite{deutsch1963attention} proposed that all stimuli are processed meaningfully before selection occurs. More recently, Lavie Load Theory (1995) \cite{lavie1995perceptual} integrated elements of both early and late selection, putting forward that attentional selection depends on the perceptual load of the task.

Addressing the limitations of selective attention is essential for several reasons:

\begin{itemize}
\item \textbf{Enhanced Safety:} In high-stakes environments such as healthcare, aviation, and security, missing critical information can have serious consequences.
\item \textbf{Improved Decision-Making:} Comprehensive awareness of the environment leads to better-informed decisions, which is beneficial in both professional and everyday contexts.
\item \textbf{Increased Efficiency:} Reducing the cognitive load on individuals allows them to focus on their primary tasks effectively, improving overall productivity.
\end{itemize}

Building on the insights from the theories of selective attention, HUX AI was developed to enhance human awareness using real-time human-AI collaboration. The HUX AI observes objects of interest and events of interest in the environment relevant to the task at hand, while the human focuses on their selective attention. The HUX AI provides additional environmental awareness by acting as an auxiliary observer, leading to better-informed real-time decision-making whenever needed. The HUX AI is context-aware, determining the information's relevance and importance. Humans can have conversations with AI based on other phenomena in the environment. Earlier, this was impossible because humans could not contextually record real-time data. However, it is now possible to use the HUX AI since it has real-time contextual data processing and understanding of the scenes.

\subsection{Video Processing}
Our work uses events of interest (EOIs) based frames for real-time video processing. The current methods for processing videos, OpenAI \cite{OpenAI_GPT4O_2024} approach with GPT-4o and GPT-4o mini utilizes a fixed sampling rate, such as extracting frames at regular intervals (such as one frame per second), which are not synchronized with event occurrences. In our work, we leverage deep learning models such as YOLO \cite{wang2023yolov7} to analyze a real-time video and extract EOIs using objects of interest (OOIs). This method enhances the signal-to-noise ratio, deriving meaningful signals from the same raw data. This reduces the computational cost due to the vision language model (VLM), thereby increasing efficiency. 

\subsection{Visual Enhancement in Extended Reality}

Visual enhancement is achieved through holographic labeling using AR algorithms \cite{park2021review} and deep learning-based algorithms. In our implementation, models such as YOLO \cite{wang2023yolov7} play a crucial role in extracting EOIs in real-time video analysis. The same models can be used for OOIs labeling in the scene, thus offering dual merits. The works of \cite{farasin2020real} present an MR system that implements real-time object detection and tracking using Microsoft HoloLens. The authors argue that this capability enhances extended reality experiences by allowing users to see holographic labels superimposed on detected objects in their environment. Our research extends this concept by feeding the enhanced scene perception to a VLM (feeding the usual scene with task-specific OOI labeling). Due to this, the HUX AI sees what the user sees in a task-specific scenario. The VLM used was not fine-tuned for this purpose. 

\subsection{Embodied AI}

\subsubsection{HCI}
Embodied AI systems such as Humane Pin \cite{humane2024} and Rabbit R1 \cite{rabbit2024} utilize VLMs to enhance HCI. These systems leverage the capabilities of VLMs to interpret and interact with dynamic scenes, providing a intuitive and responsive user experience.

\subsubsection{Robot}
Our work uses VLMs to understand dynamic scenes in HCI environments. This is similar to the approach demonstrated in AutoRT \cite{ahn2024autort}, where VLMs are used for scene understanding and grounding in robotics.

\subsection{Multi-modal HCI}
Recent advancements in gaze-facilitated information-querying systems have opened new avenues for natural and intuitive HCI. In the works of \cite{wang2024g}, the authors introduced G-VOILA. This paradigm combines users' gaze data, visual field, and voice-based natural language queries to facilitate intuitive information retrieval in daily scenarios. Their work revealed user query behavior patterns, including the prevalence of ambiguous expressions and specific eye movement patterns associated with querying tasks. 

Our HUX AI implementation extends the concept of multi-modal querying into a dynamic and real-time task-specific environment. While G-VOILA primarily focuses on static scene understanding, our system incorporates active change detection, real-time environmental adaptation, and personalized context memory. This approach addresses some identified challenges, particularly in resolving query ambiguities and adapting to users' situational context. By integrating these features, our work contributes to the evolving landscape of multi-modal, gaze-enhanced information retrieval systems, offering new possibilities for responsive and context-aware HCEI. The current implementation described in the work \cite{wang2024g} is not real-time. It also made some design choices to facilitate that potential. In the works of \cite{wang2024g}, there were four rounds of questioning where each round had one question that was asked to the AI. This contrasts with our work, where we ask several questions sequentially. 

We also extend the research by deploying this in a real-time test environment. This environment undergoes continuous changes, allowing us to test the system dynamically. We observe the system in real-time with a user who asks multiple questions. These questions involve different modalities within the same environment. This approach simulates how a person wearing smart glasses with HUX AI might interact with a real-time environment in the future. We are checking the efficacy of our system to track the multi-modal data like scene changes, and gaze changes spatially and temporally, as well as to track the context-rich user interactions.

\section{Problem Statement}
\label{sec:problem_state}
\subsection{Human Computer Environment Interaction}

The PSDs have yet to be able to aid humans in real-world scenarios. On top of their digital expertise, the next generation of devices should also help in human environment interaction, navigation, and complex domain-specific tasks users may encounter in daily life. The goal is to boost the productivity of tasks in real-world environments. The tools for domain-specific scenarios are often confined to that domain and that device. The software can not be used in other scenarios and thus can not be used as a PSD. We aim to create a PSD that enhances user productivity, decision-making, understanding, and task-specific scenarios in diverse environments. The future utility of a general-purpose AI is decided by the number of domain-specific tasks in which human-AI collaboration is possible and the magnitude of the usefulness of this AI in a task-specific domain. To achieve widespread adoption, wearables must enhance user capabilities across various daily tasks, thus functioning as a companion expert. The agenda of the HUX AI is shown in Table \ref{tab:HUX_AI_utility}.

\begin{table}[h]
\centering
    \small 
    \caption{HUX AI Agenda}
    \label{tab:HUX_AI_utility}
    \begin{tabular}{|p{0.5cm}|>{\raggedright\arraybackslash}p{3.5cm}|>{\raggedright\arraybackslash}p{3.5cm}|}
        \hline
        S. No & AI’s Agenda in Wearables & Ensures\\
        \hline
        1 & Expertise in a primary task  & High utility in at least one task, promoting regular use\\
        \hline
        2 & Proficiency in multiple daily tasks & Broad applicability throughout the day\\
        \hline
    \end{tabular}
\end{table}

\subsection{AI Utility}

\begin{figure}
    \centering
    \includegraphics[scale=0.3]{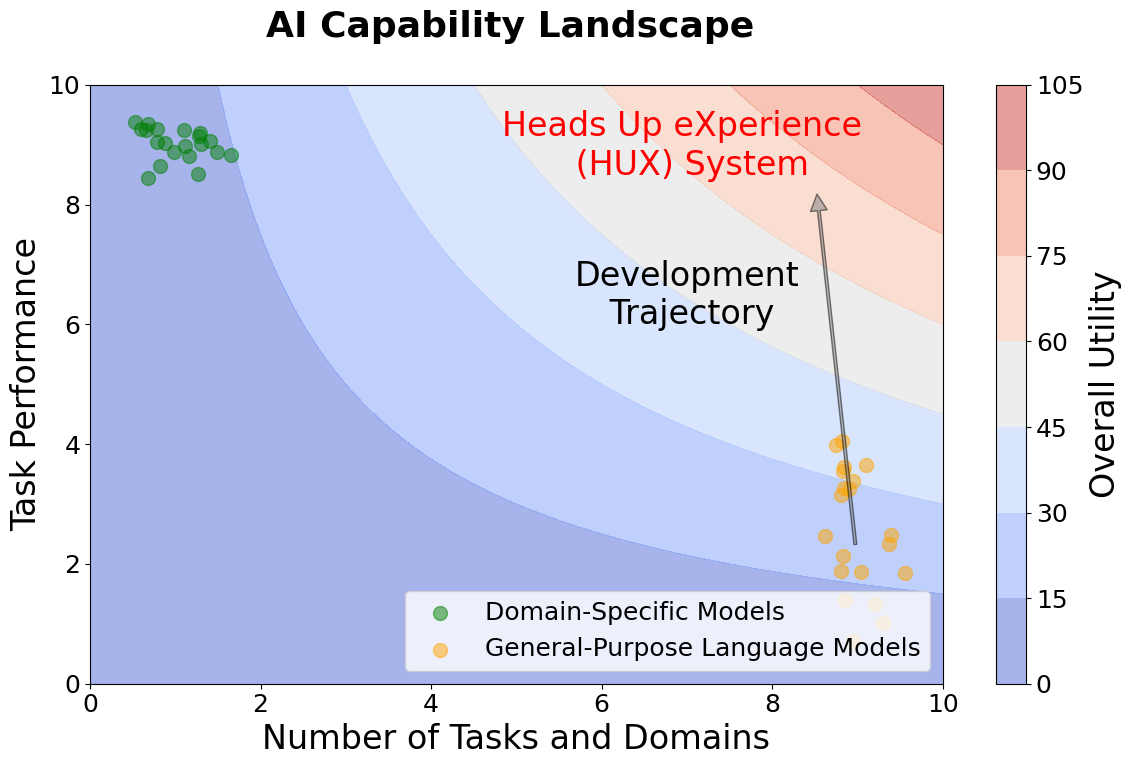}
    \caption{It shows the capability of the AI-enabled devices. A domain-specific model optimized for precise, specialized tasks (indicated in green). A generalist model offering broad, adaptable language understanding and generation (indicated in orange). HUX AI integrates a generalist base with modular, task-specific layers for flexible and specialized applications (written in red).}
    \label{fig:models}
\end{figure}

The capability of the AI-enabled devices is shown in Figure \ref{fig:models}. The domain-specific model is optimized for precise and specialized tasks indicated in green dots. An example can be the autopilot system in an autonomous car. A generalist model offers a broad, adaptable language understanding and generation, represented in orange dots. An LLM-based chatbot is an example of this type. Our HUX AI integrates a generalist base with modular and task-specific layers for flexible and specialized applications. 

AI's Utility = Number of Domains $\times$ Domain Specific Task Performance

The magnitude of the usefulness of this AI in a domain-specific task determines its necessity for that particular task. The number of domain-specific tasks in which human-AI collaboration is possible determines the usability in varied environments that the user may encounter daily, thus ensuring the usefulness of HUX AI in a PSD throughout the day. 
\subsection{Properties of a Task-Specific Environment}
The following properties are observed in a task-specific environment.

\subsubsection{Objects of Interest (OOIs):}
\begin{itemize}

\item Task-specific set of relevant objects, entities, or elements.
\item Can include physical and virtual objects (real world or AR environment).
\item May have varying levels of importance or priority within the task.
\end{itemize}

\subsubsection{Events of Interest (EOIs):}
\begin{itemize}
\item Task-specific set of relevant actions, behaviors, or phenomena.
\item Includes both expected and unexpected events due to OOIs.
\item Can involve interactions between OOIs or changes in their states.
\end{itemize}

\subsubsection{Environmental Context (EC):}
\begin{itemize}
\item Relevant background information and conditions of the task environment.
\item Influences how OOIs and EOIs are perceived and interpreted.
\item May include spatial, temporal, or situational factors.
\end{itemize}

These are the properties of a task-specific environment.

\subsection{Desired Features of AI for Varied Usability}

\begin{itemize}

\item \textit{Spatial and Temporal real-time scene understanding in a task environment:} The HUX AI tracks only EOIs, OOIs confined to a task environment. The other information it perceives has less priority, and the trackability is not guaranteed. The live computation is efficient as it avoids unnecessary spatial and temporal information. This clarifies task context and avoids mixing and overfitting data, which may overwhelm the user or the multi-modal context processing engine.

\item \textit{Understand user perception:} Generate the ROI using eye gaze data, generate enhanced visual perception using AR or deep learning algorithms for object and scene labeling, and understand them contextually.

\item \textit{Context rich interaction using multi-modal data:} Help the user with real-time queries, suggestions, and answers.

\item \textit{Understand, remember, and recall the multi-modal interaction:} Recall contextual and event information and patterns in an interaction session.

\item \textit{Future scope:} Multi-modal outputs and actions.
\end{itemize}

\section{HUX: Heads Up eXperience System and its Components}
\label{sec:hux_ai}
\subsection{The Heads Up eXperience} 
The user perception is integrated and optimized for human-AI collaboration to handle real-world or digital environments. This approach narrows the lines between HCI and human environment interaction, resulting in HCEI. The name "Heads Up eXperience" (HUX) for the AI integrated into smart glasses is inspired by the concept of a "Heads Up Display" (HUD). While a traditional HUD projects information into the user's field of view to keep them focused on their primary task, the "Heads Up eXperience" extends this concept by emphasizing a immersive, interactive, and multi-modal interface.

The term "experience" reflects the capability of HUX AI to not only present relevant information but also enhance user engagement with their surroundings in a dynamic and contextually aware manner. By integrating various functionalities typically managed by separate devices into the smart glasses, HUX AI could eliminate the need for smartphones and other gadgets. This consolidation allows users to perform a wide range of actions, access information, and interact with their environment directly through the heads-up interface, streamlining the user experience and facilitating a intuitive interaction with technology.

\subsection{The HUX AI Architecture}
\begin{figure*}
    \centering
    \includegraphics[width=1\linewidth]{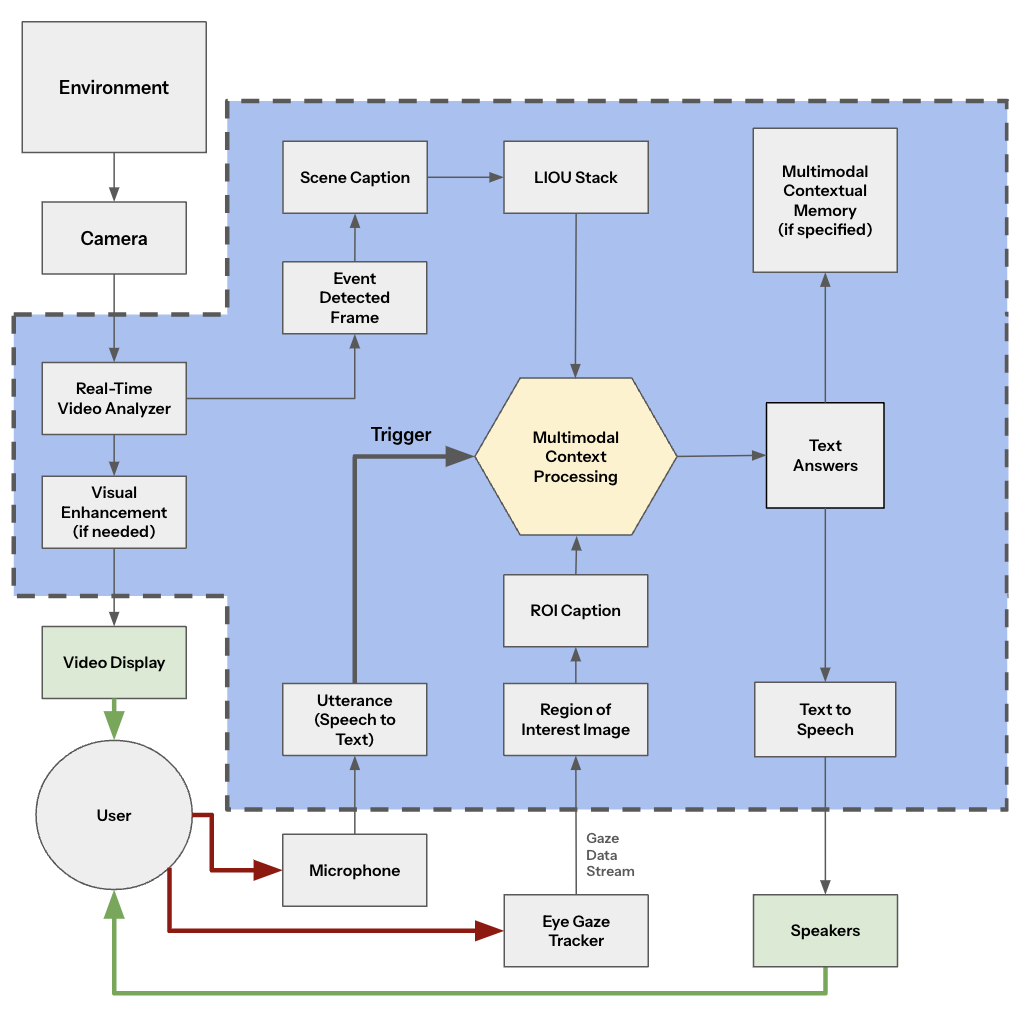}
    \caption{HUX AI Architecture: The origin of multi-modal data, data processing, multi-modal context processing, the generation of outputs. The "LIOU Stack" means Last-In-Only-Used Stack"}
    \label{fig:HUXAI_Archi}
\end{figure*}

Figure \ref{fig:HUXAI_Archi} shows all the integral components for the system's working and flow. 
\begin{itemize}
\item  It shows the sources of all data modes, the data flow inside the HUX AI system, data processing, and the data flow for each modality.
\item  All input data modes are converted to text captions and descriptions and are integrated by the multi-modal context processing engine using a prompt architecture.
\item  Creation of a new voice file (new utterance input) triggers a multi-modal context processing episode.
\item  Collects the latest scene captions stored in the Last-In-Only-Used (LIOU) stack.
\item  Gets the latest ROI caption whose origin image is created simultaneously with the speech audio file.
\end{itemize}

Multi-modal Context Processing receives the textual representations of all the possible input modalities, processes the data, and returns the output to the user. The environment here is dynamic (subject to change), and the user's goal is to accomplish the task of interest in this environment. The boxes highlighted in green indicate the processed perception input given to the user by the HUX AI system. The red arrow indicates the inputs given by the user to the HUX AI system. The Green arrow indicates the outputs processed or generated by the HUX AI system for the user. The grey arrow represents the trigger that initiates a context processing instance. Other arrows represent the data flow from one block to the other. The blue shaded area indicates the HUX AI system, and the rest are not a part of the HUX AI system.

\subsubsection{Models used}
The core models that were used are as follows:
\begin{itemize}
\item \textbf{Vision Language Model:} \textit{Mini Intern VL 1.5 2 Billion Parameter}, \cite{chen2024far} \cite{chen2023internvl} was used for image captioning.

\item \textbf{Large Language Model:} \textit{Llama-3 8 Billion Parameter model} \cite{meta_llama3} was used for all LLM instances, including the multi-modal Context Process

\end{itemize}

The models for task-specific perception vary based on the task. Some models used are YOLO \cite{wang2023yolov7} and printed circuit board defect detection \cite{test2024objectdetection} for identifying OOIs and EOIs.

\subsubsection{Core Components}
The various modalities used are as follows: 
\begin{itemize}
\item {\texttt{User's Speech Input}}: Listen to user speech inputs, which may be instructions, queries, or general conversations. Trigger a multi-modal context processing instance.

\item {\texttt{Eye Gaze Information}}: User eye gaze is measured and ROI is produced.

\item {\texttt{Real-time Video Analyser}}: Real-time task-specific EOI and OOI detection is done in the environment scene, and frames are extracted. 

\item {\texttt{Vision Language Model}}: To convert input image to image captions whenever needed.

\item {\texttt{Large Language Model}}: Used primarily as a multi-modal data context processing engine. It is also used when an LLM instance is needed.

\item {\texttt{Multi-modal Contextual Memory}}: It is a data structure that stores context-rich multi-modal data for long-term memory retrieval.
\end{itemize}


\subsection{Real-Time Video Feed Filtering based on Events}

\begin{figure*}
    \centering
    \includegraphics[width=1\linewidth]{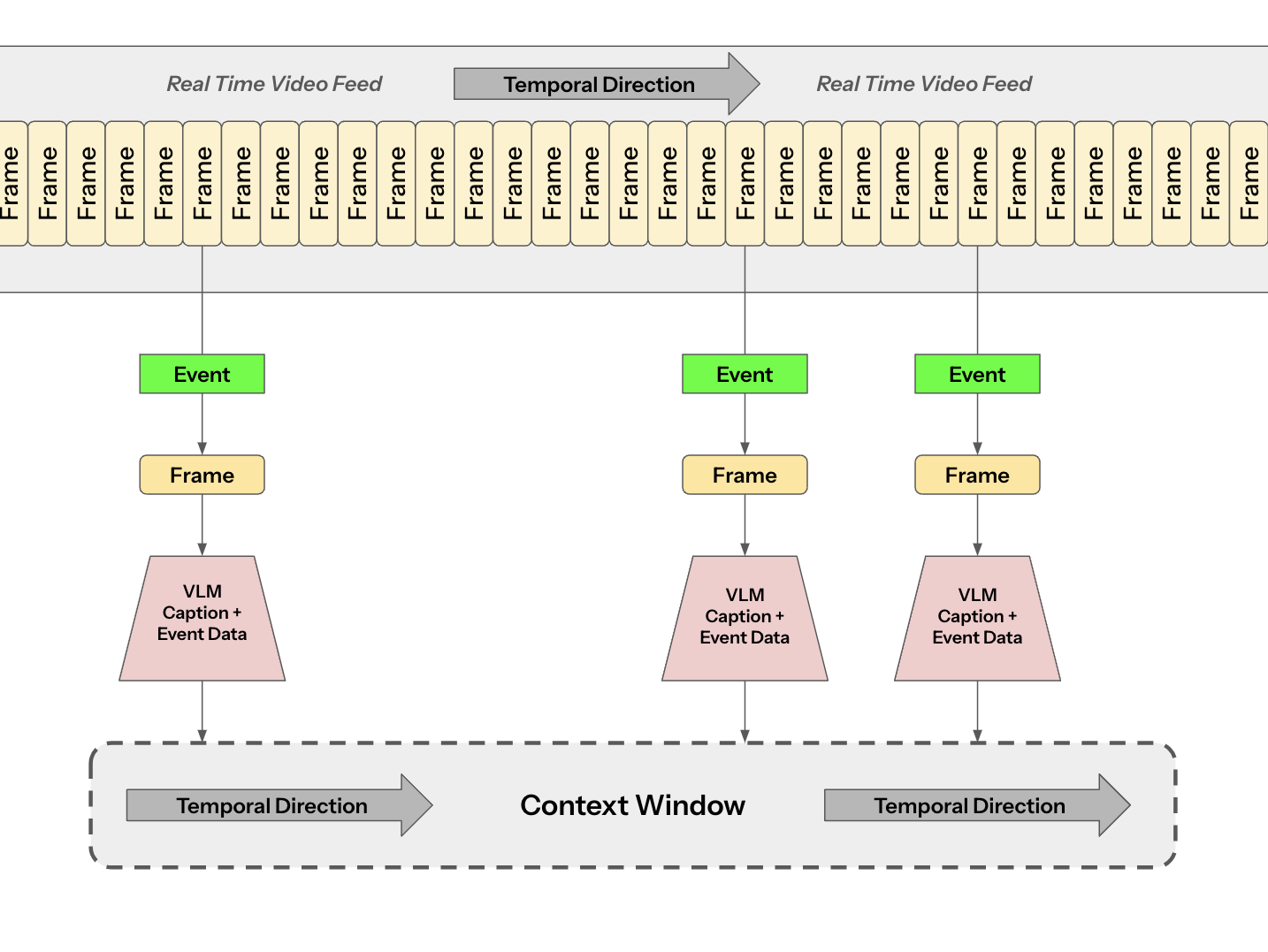}
    \caption{Expected: real-time Event-based Video Context Processing using VLMs for detected events of interest.}
    \label{fig:video_filter}
\end{figure*}

Figure \ref{fig:video_filter} represents the temporal flow of frames from a real-time video. The event indicated by the green boxes indicates the filtered EOIs found by the real-time video analyzer. These EOI frames are sent to the VLM for image captioning. The event data derived from the real-time video analyser are combined with image captions and sent sequentially for context processing with metadata. The data from selected frames triggered by certain events are updated in the context window. 

The EOIs can be obtained from real-time video of the environment:
\begin{itemize}
\item \textit{Change in the number of objects of interest within the frame:} This includes the appearance or disappearance of specific objects or a total number of objects.

\item \textit{Change in the object behavior:} This also includes rapid changes in bounding box sizes, swift movements across the frame, changes in the scale or proximity of objects, and other feature variations.

\item \textit{Custom-defined events specific to the application domain:} This involves detecting custom-defined events such as gestures, actions, or environmental changes.
\end{itemize}

Later, these EOIs are exported as frames to be processed by the VLM, and the event data from the real-time video analyzer to be processed by the context processing engine to reconstruct the temporal scene textually. This way, the context-processing engine understands the occurrences in real-time while saving significant computing and processing time. 

\begin{figure*}
    \centering
    \includegraphics[width=1\linewidth]{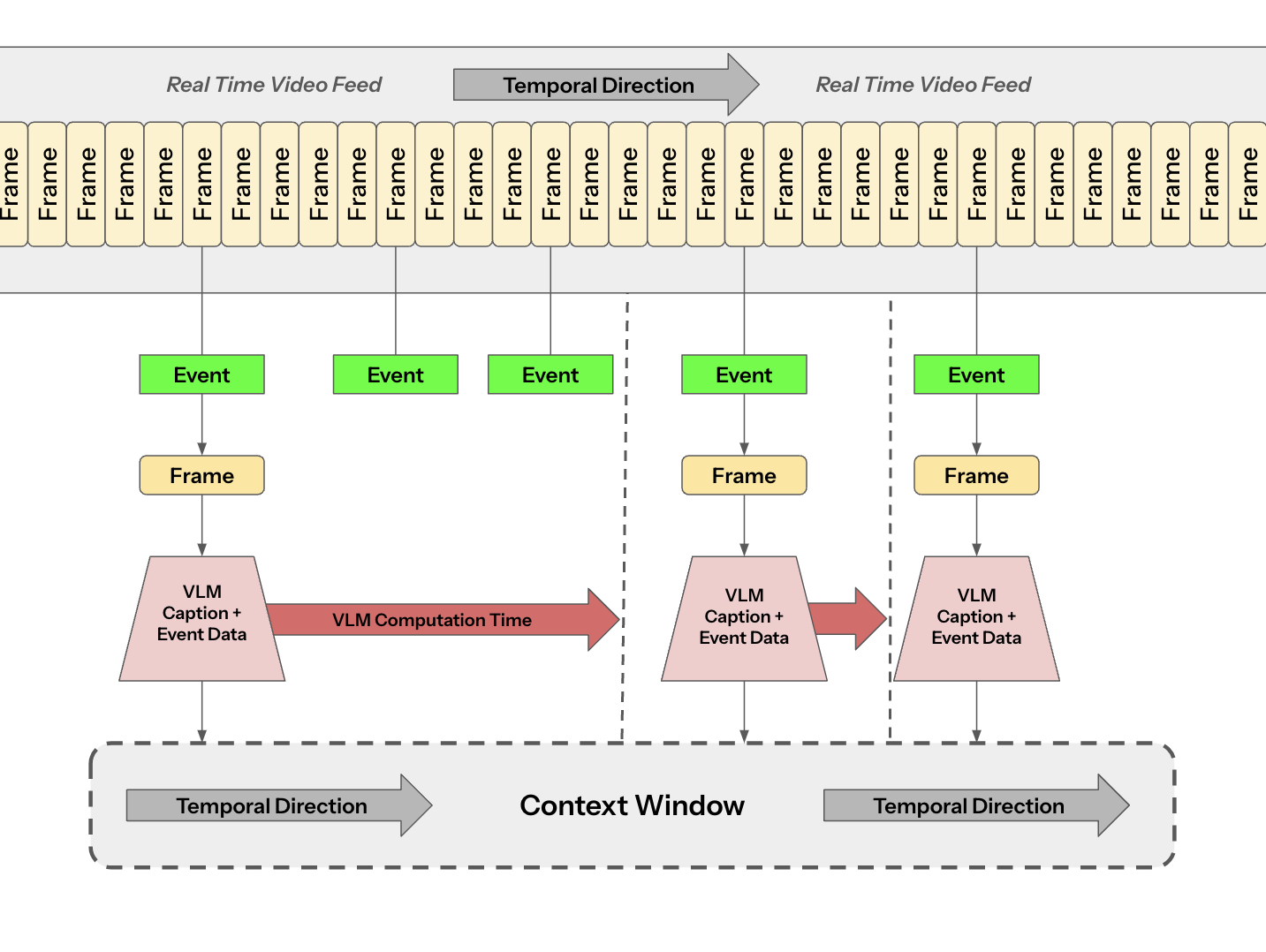}
    \caption{Reality: real-time event-based Video Context Processing using VLMs for detected events of interest. Some events are missed due to the computation time of the previous frame. The arrow "VLM Computation Time" specifies that the VLM is busy processing the previous frame and that subsequent events of interest are missed until freed.
}
    \label{fig:video_filter_3}
\end{figure*}

\begin{figure*}
    \centering
    \includegraphics[width=1\linewidth]{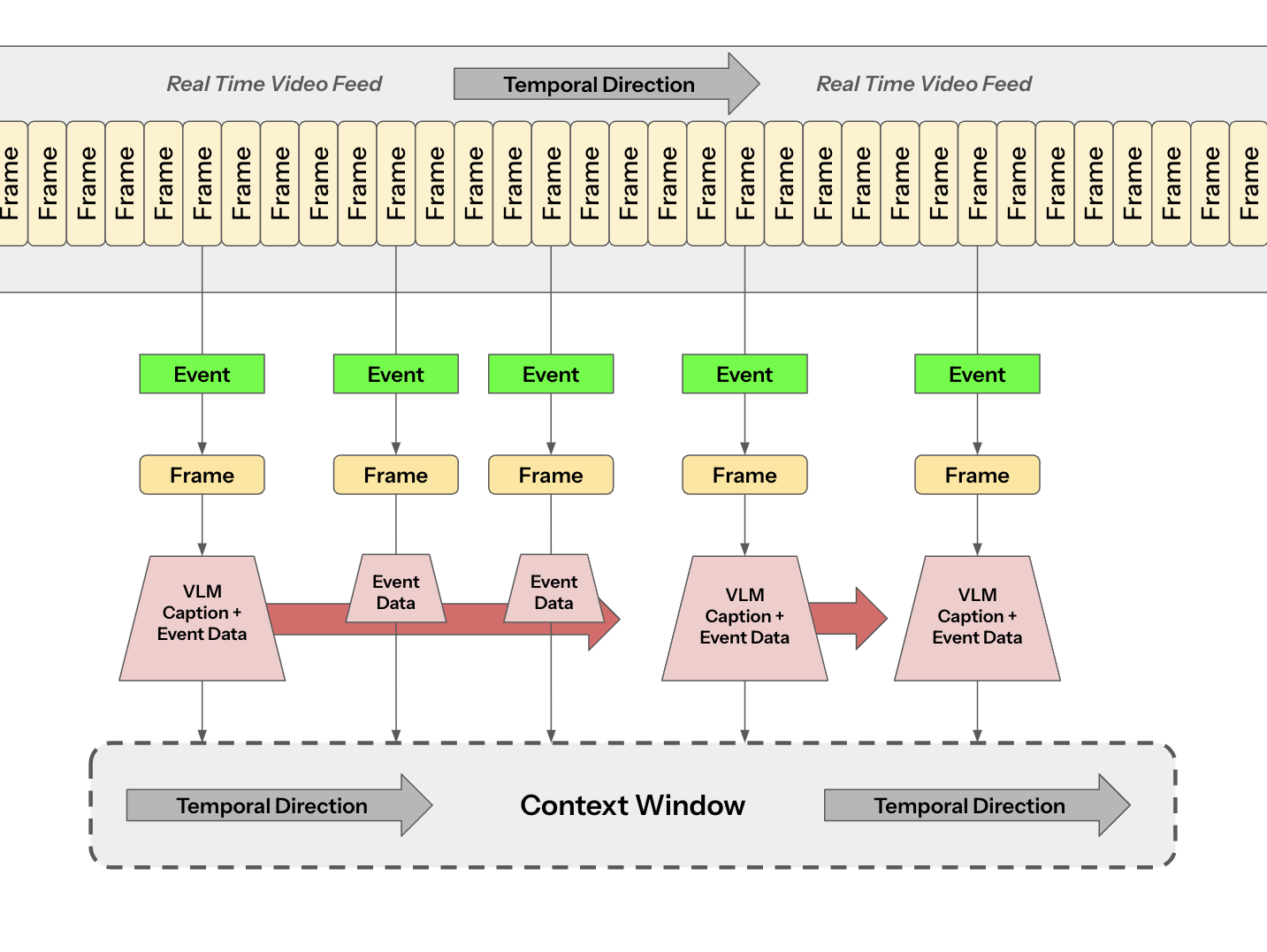}
    \caption{A Hybrid approach for real-time event-based Video Context Processing using VLMs to extract information from all the detected events of interest.}
    \label{fig:video_filter_2}
\end{figure*}

There is a possibility that during the VLMs processing of a frame, more than one frame might get classified as an EOI. After the VLM completes processing the current frame, the middle frame is skipped over to focus on the latest frame, as depicted in Figure \ref{fig:video_filter_3}. By having a hybrid approach as depicted in Figure \ref{fig:video_filter_2}, the real-time video analyzer can give event data (as essential as the counts of the objects), which can help us track the contents even in the absence of detailed image captioning which finds complex relationships and descriptions of the scene. This problem occurs only when there is a need for real-time computation.

\subsection{Task Specific Scene Processing}
The above section showed the merits of using a real-time scene analyzer to identify EOIs. We use the same real-time video analyzer to enhance user perception visually.

\subsubsection{Task Specific Scene Enhancement}
The same computer vision model used above can be re-purposed here to temporally filter out events and visually annotate the scene image feed with OOIs confined to the user task. 

The merits of this approach are as follows:
\begin{itemize}
\item Visual aid for the user if the user's scene image feed is annotated when working in a task-specific scenario to improve task-specific selective attention.

\item  HUX AI is aware of the scope of task-specific OOIs among other objects.

\item  Modularity to handle different tasks with a unique set of OOI and EOI. 

\item Adaptive attention allocation: Dynamic shifting of focus based on changing environmental priorities.

\item  Attention restoration: Improved techniques for mental recovery and focus renewal in the scene.

\item  Selective ignorance cultivation: Helping the user to disregard non-essential information for the task intentionally.
\end{itemize}

\subsubsection{Modularity of Task Specificity}
Switching of task-specific computer vision models responsible for detecting OOIs can be executed using agentic behavior \cite{Liu_LlamaIndex_2022}. The switching can be done with or without explicit user summons. Since the agent can understand the context of the initial scene, this insight can be used to decide if it has to use task-specific computer vision models (defined as tools) for further HCEI.

\begin{figure*}
    \centering
    \includegraphics[width=1\linewidth]{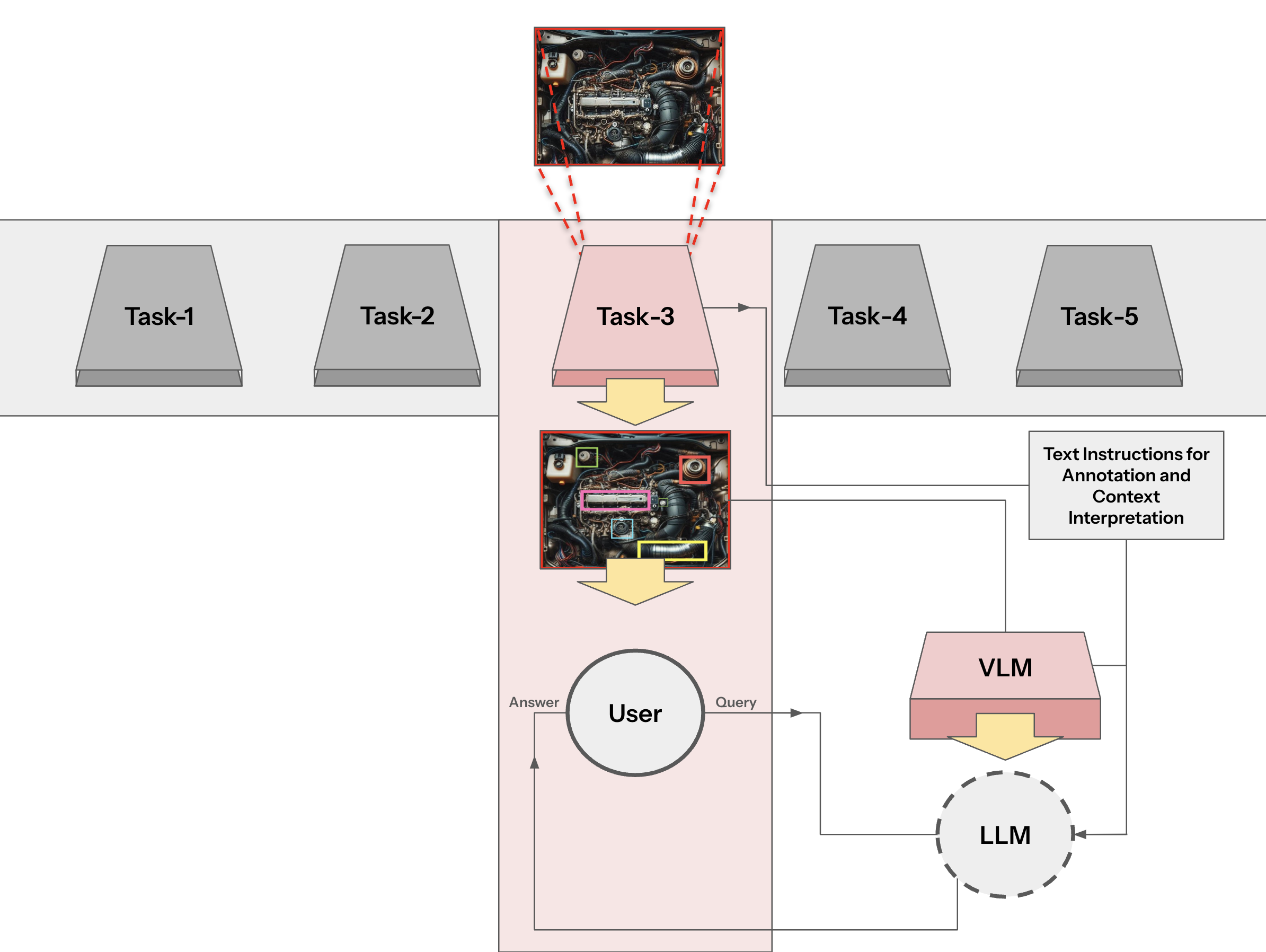}
    \caption{Task Specific Scene Enhancement, Question and Answering. Task-1, Task-2,... Task 5 indicates the different computer vision models.}
    \label{fig:task_specific}
\end{figure*}

Figure \ref{fig:task_specific} depicts task-specific computer vision models and other components. The different tasks Task-1, Task-2,... Task 5 indicates the different Computer Vision models that perform EOI detection and OOI detection in task-specific scenarios that a user might encounter. The user visual scene is enhanced by the computer vision model aligned with the context of the chosen task. They visually label the different OOIs in the scene, acting as a visual aid for the user. The user perceives this in a display. Figure \ref{fig:task_specific} also shows how the same image is passed to the VLM for the HUX AI system to perceive what the user perceives to understand and empathize with user perception. The text instructions and annotations for context interpretation that are unique for specific tasks are passed down to VLM (for better scene interpretability) and LLM (for better context understanding). Later, the user can have conversations with the LLM regarding the scene. Figure \ref{fig:task_specific} also represents the modularity in swapping the different task-specific models while retaining the core components.

\subsection{Eye-Gaze Processing}
\begin{figure*}
    \centering
    \includegraphics[width=1\linewidth]{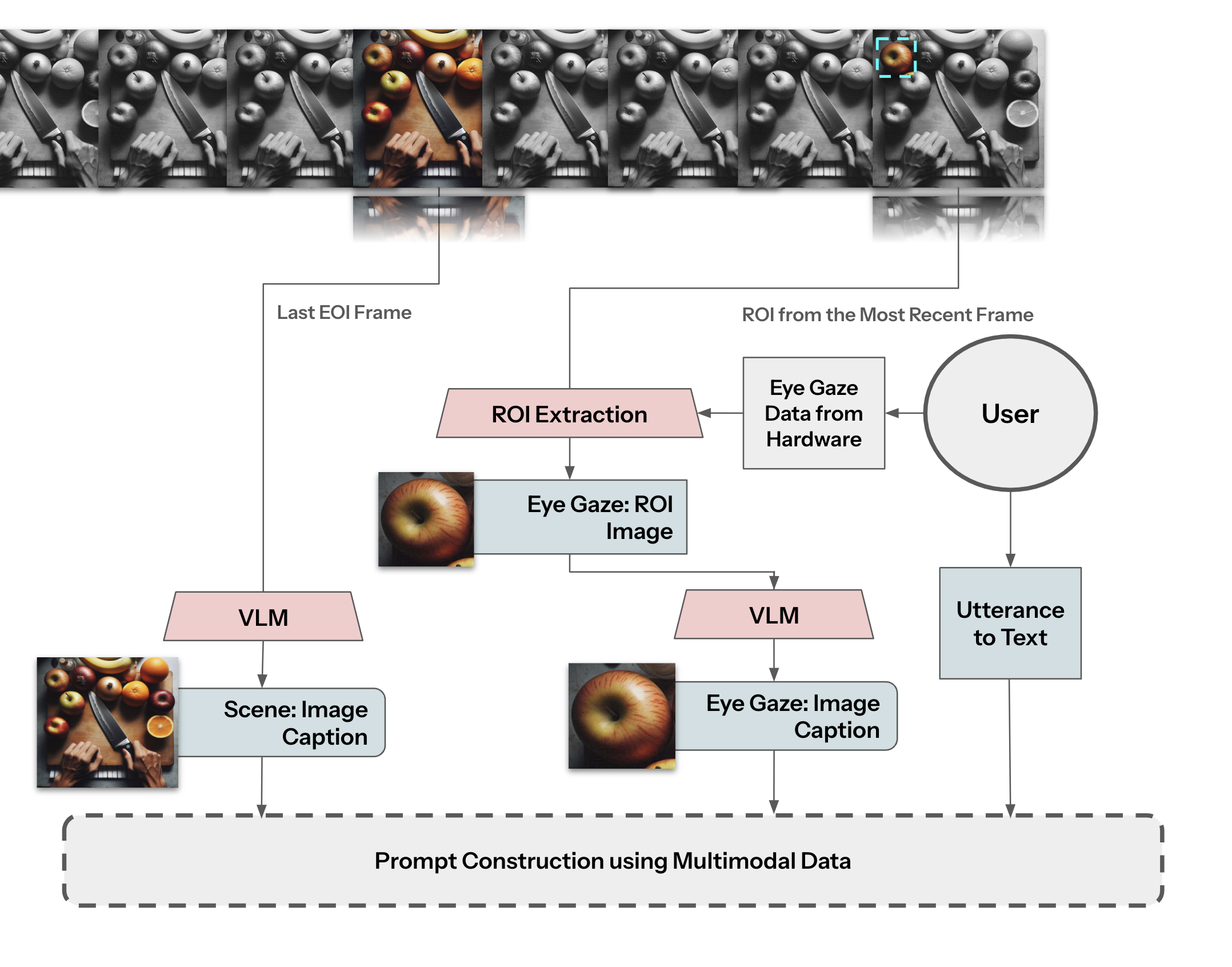}
    \caption{Processing of Eye Gaze in parallel with Scene Processing}
    \label{fig:Eye_Gaze}
\end{figure*}

This pipeline discusses how the eye gaze data is processed in real time in parallel to the scene data. Figure \ref{fig:Eye_Gaze} shows the origin and flow of the eye gaze data. The scene image processing is shown here to demonstrate the parallel and interdependent nature of the gaze and scene processing. 

The Figure \ref{fig:Eye_Gaze} is explained as follows: 
\begin{itemize}
\item Colored image indicates the frame where a new EOI is observed.
\item Subsequent black and white frames do not have a new EOI, thus not sent for VLM processing for image captioning, thus demonstrating compute efficiency.
\item The highlighted apple in a blue dotted bounding box indicates the frame at which speech was triggered. The gaze coordinates at that time instance are noted, and the scene image from the same time instance is used to generate the ROI gaze image.
\item Frame with the EOI is sent to the VLM for image captioning.
\item Extracted ROI from the scene image and eye gaze data from the hardware is sent to the VLM for image captioning.
\item  Utterance is derived from the user and converted to text using OpenAI Whisper.
\item  All three input data are inserted into the prompt for spatial and temporal scene recreation in text descriptions.
\end{itemize}

Eye Gaze data is obtained from the wearable hardware. The data could be gaze coordinates in (x,y) and the perimeter of the gaze area.
Raw gaze data and the overall scene are used to obtain the ROI by cropping or segmenting from the overall scene, which the VLM uses to obtain text inferences. The scene image is also passed to the VLM. The image captions are obtained from both the VLM. The data is translated into textual form and passed to the multi-modal context processing engine.

\subsection{Multi-modal Contextual Memory Creation}
\label{subsec:mul_memory}

Upon summoning, the multi-modal data and rich user context during an interaction can be stored in a structured multi-modal memory format. Later, it can be used for retrieval based on mere contextual cues using Retrieval-Augmented Generation (RAG) \cite{lewis2020retrieval} rather than explicitly searching for one of the data modes like images, information, or conversation context. The Llama Index \cite{Liu_LlamaIndex_2022} was used to execute this specific pipeline. Agentic behavior was utilized to extract the class of object of interest from naive user input interaction. 

\begin{figure*}
    \centering
    \includegraphics[width=1\linewidth]{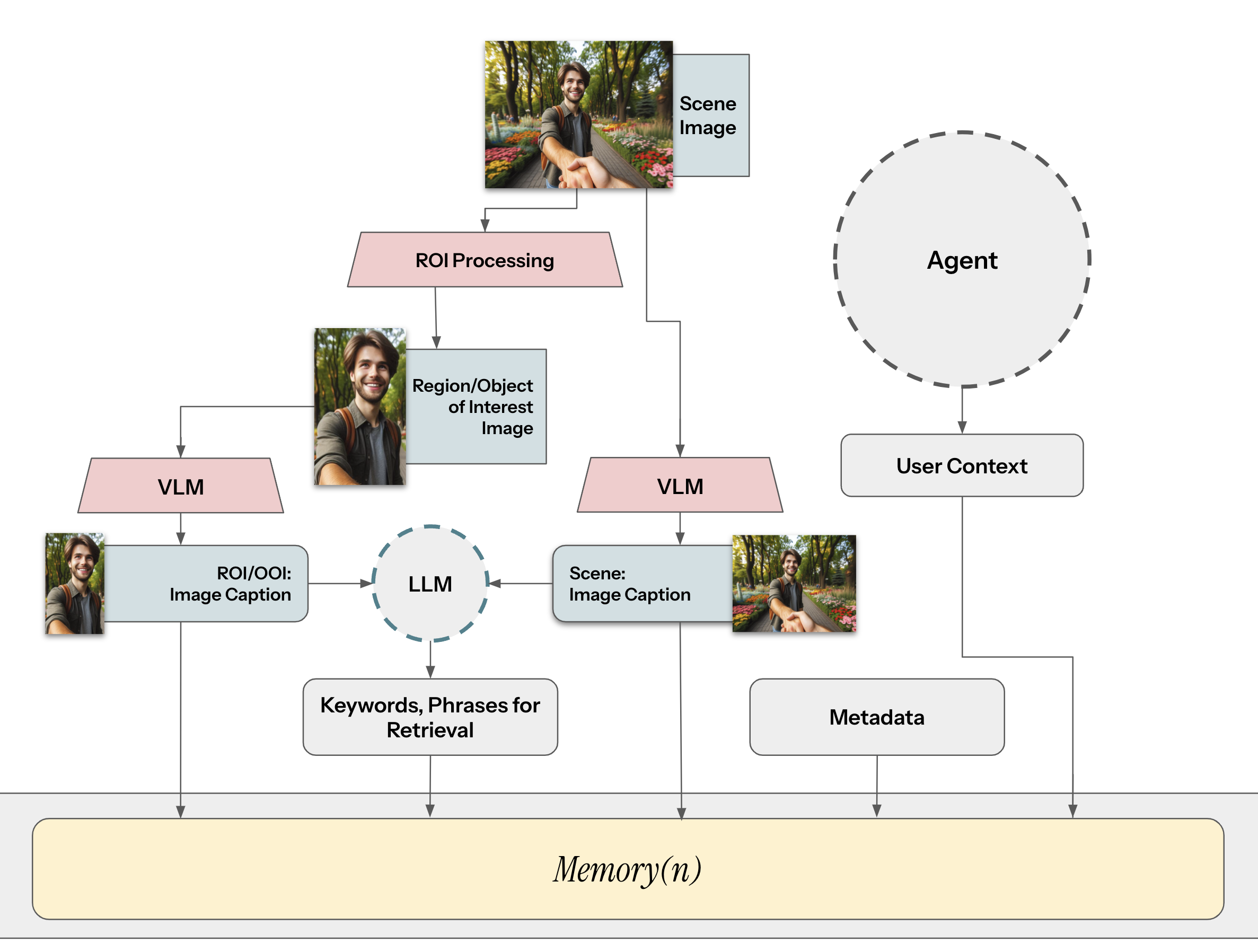}
    \caption{Process of multi-modal Contextual Memory Pipeline}
    \label{fig:Memory_Pipeline_1}
\end{figure*}

Figure \ref{fig:Memory_Pipeline_1} shows the process flow of how it can be implemented. The chosen scene image is sent to the VLM for image captioning. The same scene image is sent for ROI processing and extracting. ROI can be processed using multiple inputs, such as verbal specifications of an OOI or eye gaze ROI. This ROI/OOI is sent to VLM to get image captions. The agent extracts the user context from the conversation. All the multi-modal data is sent to an LLM instance to generate keywords, phrases, and other contextual cues can be retrieved in later time period by tools like RAG \cite{lewis2020retrieval}.

\begin{figure}
    \centering
    \includegraphics[scale=0.2]{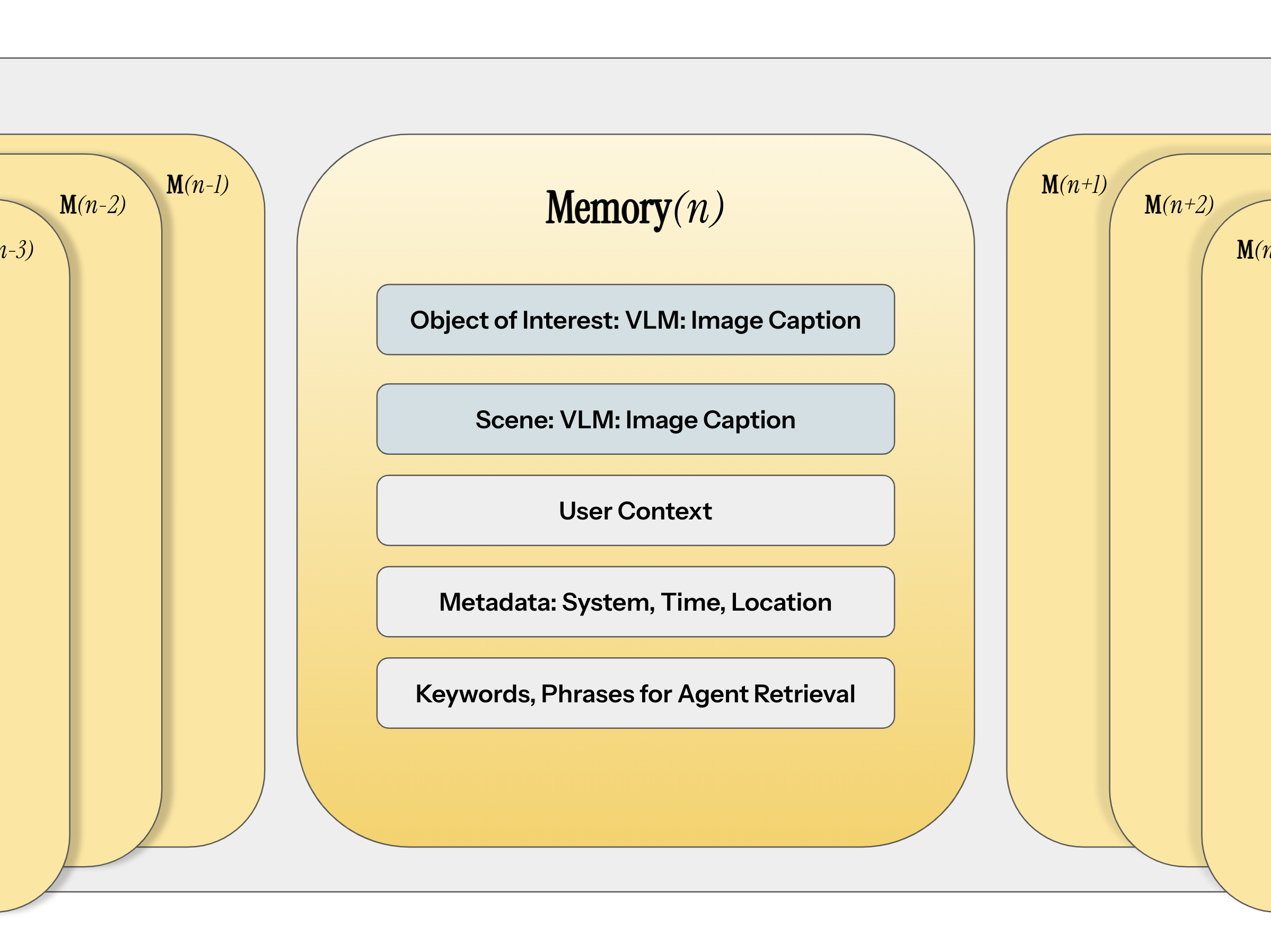}
    \caption{Data Structure of a multi-modal Contextual Memory}
    \label{fig:Memory_Pipeline_2}
\end{figure}

Figure \ref{fig:Memory_Pipeline_2} shows the data structure of a multi-modal contextual memory and how it exists among other memories. The following describes each data type in the multi-modal memory data structure.

\begin{itemize}
\item {\texttt{Object of Interest}}: The YOLO \cite{wang2023yolov7} was used to obtain the class OOI image from the scene image. Image Captions of Object/Region of Interest + Object/Region of Interest location are stored here.

\item {\texttt{Scene}}: Image Captions of Scene + Scene Image Location is stored here.

\item {\texttt{Speech Input}}: Context given by the user during an interaction about this object, like its relationship with time, people, other objects, and emotions, is obtained from the conversation.

\item {\texttt{Meta Information}}: Metadata like location, time, and device name are obtained.

\item {\texttt{LLM Generated Content}}: Keywords, phrases generated by an LLM instance for future retrieval. The advantages of this are that it minimizes the computation cost in the future when RAG \cite{lewis2020retrieval} tries to process the context of all raw data to match the user's query. By having a set of probable user queries (generated by the LLM instance) that could be asked by the user in the future to refer to this memory, the time and computation required to find the data can be reduced. More work has to be done to explore the efficacy.
\end{itemize}

Later, when the user refers to the object with the subtlest of the direct or indirect contextual cues, the agent can refer to the stored memories to retrieve the source image, data, and context.

\section{HUX AI Applications}
\label{sec:applications}
\subsection{Solving Selective Attention Problem in real-time Human Environment Interaction}
The merits of solving the selective attention problem are as follows: 

\begin{itemize}
\item The goal of AI is to pay attention to tasks outside the scope of human selective attention.

\item \textit{Enhanced situational awareness:} Improved detection of subtle environmental cues or threats in complex scenarios.

\item \textit{Cognitive offloading:} By monitoring less salient task elements, the HUX AI could free up human cognitive resources for higher-level decision-making.

\item \textit{Enhanced multitasking support:} In complex scenarios requiring attention to multiple sub-tasks, the AI could manage parallel information streams more effectively than humans.
\end{itemize}

\subsection{Spatial and Temporal Scene Recall}
There are two scenes: Scene 1, with five demo videos of 30 seconds each, and Scene 2, with one demo video of 13 seconds. The scenes had several fruits placed in a dark background. A few of the fruits were removed from the scene and were replaced in the background several times. The YouTube link for the video is here: (https://youtu.be/KyQnhxZyu08). The demo inference of scene 1, demo video one, is as follows: 

\textit{In comparing HUX AI computer vision analysis with manual video analysis, both methods generally align but show some discrepancies. HUX AI identifies the highest variety of fruits at timestamps $20240708_163548$ and $20240708_163556$, while manual analysis identifies it at $20240708_163600$. Both agree that there is no consistent pattern in adding or removing fruits and that apples appear in 7 frames, disappearing once. HUX AI states that no consecutive frames have the same arrangement, while manual analysis disagrees. This is because no frame was processed, and no EOI was detected. Both agree that the orange has the most fragmented presence and that bananas dominate the frequency and quantity index. The banana is noted as the only consistently present fruit throughout the video.} 

\textbf{Selective real-time data computation:} All data can be stored without real-time computing. It can be computed when necessary. Only the frames with EOIs are computed in real time. This is because real-time computation must occur selectively due to resource and time constraints.  

\subsection{Task Specific Scene}
Demonstrating a multi-modal task-specific scene perception: Task-specific visual enhancement using AR algorithms and deep learning computer vision algorithms. VLM is used in the image captioning of this enhanced scene.

\begin{figure}
    \centering
    \includegraphics[scale=0.25]{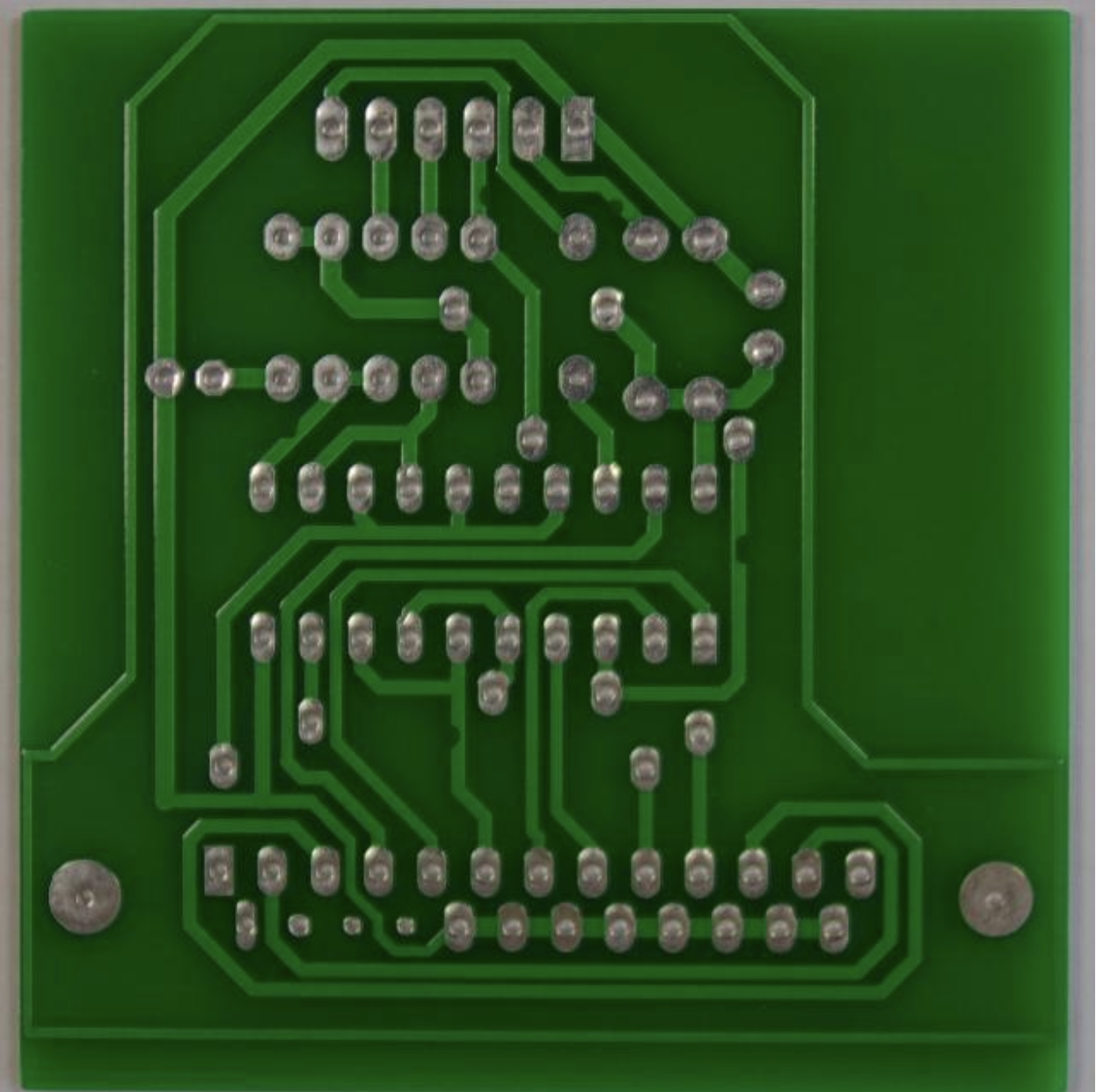}
        \includegraphics[scale=0.23]{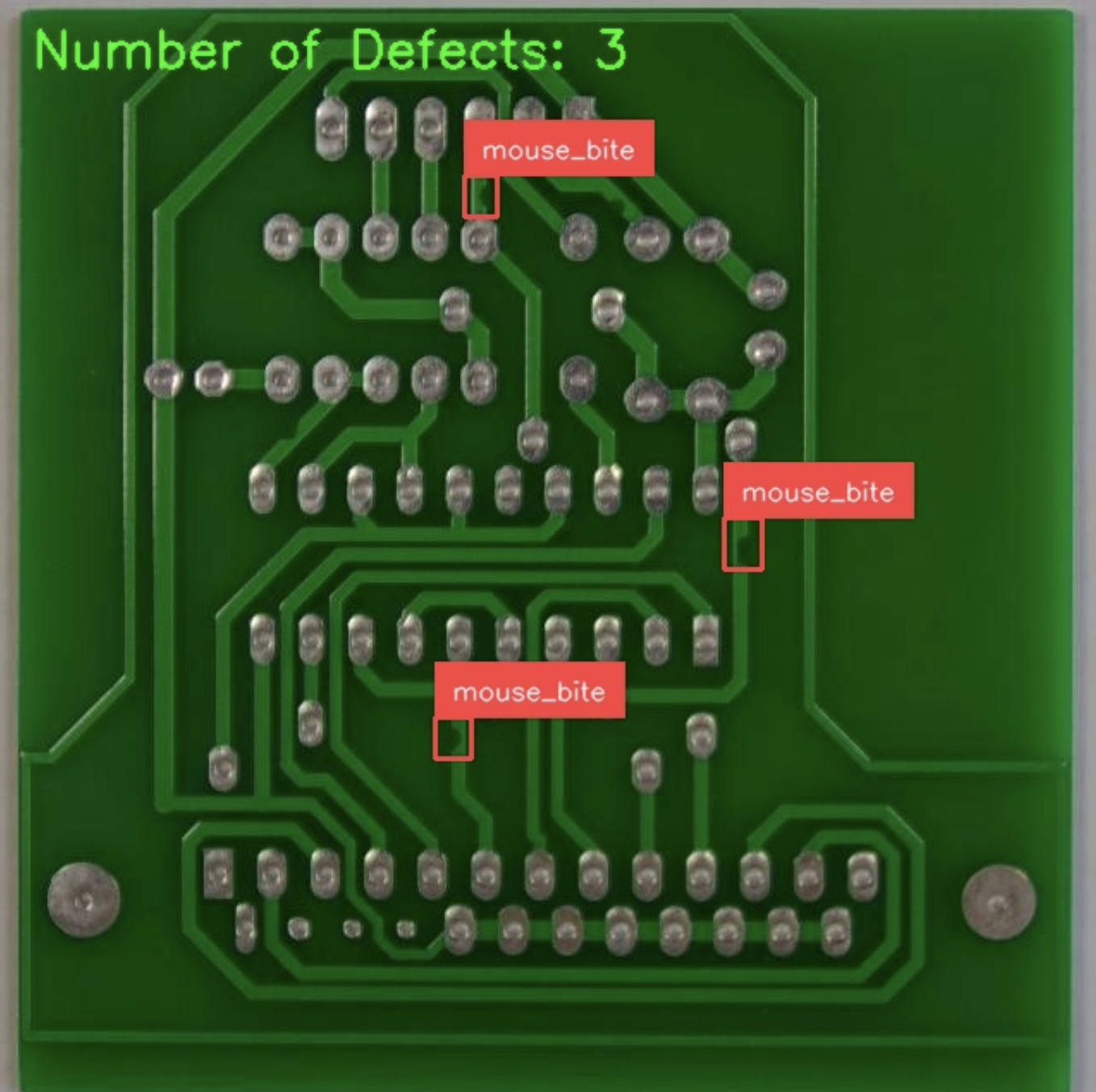}
    \caption{Left: PCB in the unlabelled condition. Right: PCB labeled with a computer vision model whose only goal is to find the defects in the PCB}
    \label{fig:Unlabelled_PCB}
\end{figure}

\textbf{Experiment Scenario:} An Engineer looking at a PCB, shown in Figure \ref{fig:Unlabelled_PCB}.

The Figure \ref{fig:Unlabelled_PCB} (left) in \cite{object-detection-ycict_dataset} PCB dataset shows the unlabelled image of a PCB that may have a different number of features. The goal is to find the defect in the PCB using the computer vision model \cite{test2024objectdetection} and identify the possible defects in the PCB circuit among six classes. The image was labeled using the same computer vision model (Figure \ref{fig:Unlabelled_PCB} (right)).

We leverage LLM-based agentic behavior \cite{Liu_LlamaIndex_2022} to automate the usage of this model only when required. It is triggered automatically based on user perception instead of the user explicitly mentioning the use of this model.

\subsubsection{Demonstrated Workflow}
The workflow is shown in Table \ref{table:user_agent_interactions}. It is briefly described here:

\textbf{User Requests Help with Visual Identification:}
\textit{Action:} User asks for help understanding what they are seeing.
\textit{Response:} Agent visually inspects and identifies the components on the PCB.

\textbf{User Inquires About Defects:}
\textit{Action:} User asks if there are any defects in the circuit.
\textit{Response:} Agent uses the PCB-HUD tool which has the \cite{test2024objectdetection} PCB defect detection model.

\textbf{Agent Analyzes and Infers Defects:}
\textit{Process:} Agent uses VLM to perform image captioning of the labeled scene image.
\textit{Conclusion:} Agent identifies three defects (mouse bites) and describes their spatial locations verbally (usage of VLM).

\textbf{User Seeks More Information on Defects:}
\textit{Action:} User asks for an explanation of the defects.
\textit{Response:} Agent explains what a mouse bite defect is based on its knowledge. Bridging knowledge from task scene enhanced by this task-specific computer vision model and LLM's trained knowledge.

\textbf{User Checks Agent's Memory:}
\textit{Action:} User asks how many defects were seen.
\textit{Response:} Agent confirms three defects and reiterates their locations.

\FloatBarrier
\begin{table}[h]
\centering
\scriptsize 
\caption{Demo User-Agent Interactions in a Task-Specific Scenario}
\label{table:user_agent_interactions}
\begin{tabular}{|l|p{10cm}|}
\hline
\textbf{Interaction} & \textbf{Conversation} \\ \hline
1 & \textbf{User :} "What am I seeing?" \newline \textbf{Agent:} The image depicts a complex printed circuit board (PCB) with various electronic components, including resistors, capacitors, and integrated circuits (ICs). The PCB is green, typical for PCBs, and features soldered connections, wires, and joints. \\ \hline
2 & \textbf{User :} "Are there any defects in the circuit?" \newline \textbf{Agent Thought:} I need to use a tool to help me analyze the PCB. \newline \textbf{Action:} PCB \newline \textbf{Action Input:} {checkfor: 'defects'} \newline \newline Agent decides that it has to use the PCB\_Model to find defects in a circuit. \newline \newline Labelled Image is received from the PCB\_Model. \newline \textbf{VLM Input:} {'check' + checkfor + ‘use labels.'} \newline \textbf{VLM Observation:} There are three defects in the image: \newline 1. \textbf{Mouse\_bite}: This defect is located at the bottom right of the circuit board. It appears as a small, irregularly shaped area on the circuit board. \newline 2. \textbf{Mouse\_bite}: This defect is also located at the bottom right of the circuit board. It appears as a small, irregularly shaped area on the circuit board. \newline 3. \textbf{Mouse\_bite}: This defect is located at the top left of the circuit board. It appears as a small, irregularly shaped area on the circuit board. \newline The defects are marked with red boxes and labeled with "mouse\_bite" in white text. \newline \textbf{Agent:} There appear to be three defects, all of which are mouse bites located at the bottom right and top left corners of the circuit board. \\ \hline
3 & \textbf{User :} "What does this mean?" \newline \textbf{Agent Thought:} I'll explain what a mouse bite is in the context of PCB defects. \newline \textbf{Agent:} A mouse bite is a type of defect that occurs when there's an imperfection or damage to the copper foil on the circuit board. It can be caused by physical stress, such as bending or flexing, which can lead to cracking or lifting of the copper layer. Mouse bites can cause electrical connections to fail, leading to errors or malfunctions in the system. \\ \hline
4 & \textbf{User :} "How many problems did you see?" \newline \textbf{Agent Thought:} I can answer without using any more tools. \newline \textbf{Agent:} There are three defects identified as "Mouse\_bite" located at the bottom right and top left of the circuit board. \\ \hline
5 & \textbf{User :} "How many flowers did you see?" \newline \textbf{Agent Thought:} I think there might be some confusion here, we were discussing a printed circuit board (PCB) earlier, not flowers. There's no mention of flowers in our conversation so far. \newline \textbf{Agent:} None \\ \hline
\end{tabular}

\normalsize 
\end{table}
\FloatBarrier

\textbf{User Asks an Out-of-Context Question:}
\textit{Action:} User asks how many flowers were seen.
\textit{Response:} Agent identifies the question as out of context and responds appropriately, noting there were no flowers.

\subsection{Multi-modal Contextual Memory Creation}
\label{subsec:multimodalcontext}
\begin{figure}[h]
    \centering
    \includegraphics[scale=0.11]{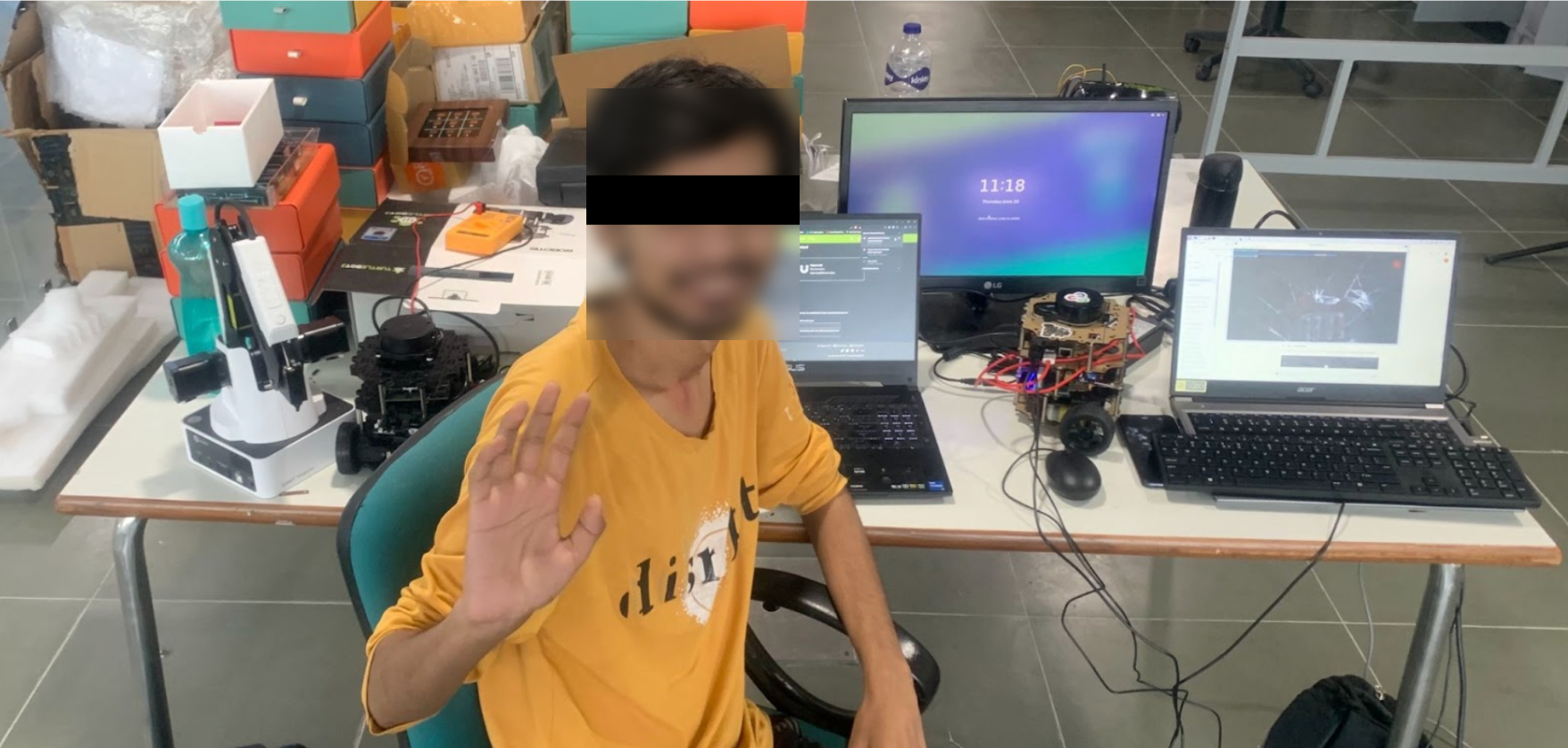}
    \includegraphics[scale=0.12]{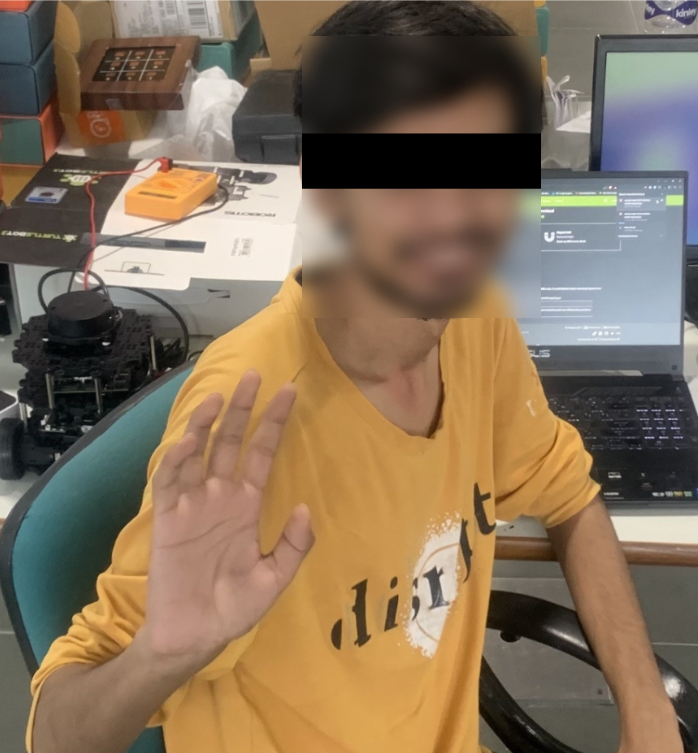}
    \caption{Left: The scene image captured by a PSD. Right: ROI extracted by the agent using conversation context}
    \label{fig:roi_psd}
\end{figure}

The Subsection \ref{subsec:mul_memory} explains the multi-modal contextual memory creation. Table \ref{tab:memory_storage} shows how the multi-modal contextual memory creation can happen in a real-life setting where the user asks the HUX AI system to remember an engineer he met for the first time with the multi-modal rich information. This is shown in Figure \ref{fig:roi_psd}.

\FloatBarrier
\begin{table*}[h]
    \centering
    \scriptsize 
        \caption{Demo of Multi-modal contextual memory storage for new engineer Imam}
    \label{tab:memory_storage}
    \begin{tabular}{|>{\raggedright\arraybackslash}p{2.5cm}|>{\raggedright\arraybackslash}p{9cm}|}
        \hline
        \textbf{Attribute} & \textbf{Details} \\
        \hline
        Object of interest & The person in the image is a young man with dark hair and a beard. He is wearing a yellow t-shirt with some text and graphics on it. He is seated in front of a computer setup, which includes a laptop, a computer monitor, and various electronic components and devices. The man is smiling and waving at the camera, giving a friendly and approachable appearance. \\
        \hline
        Scene Description & The image depicts a young man sitting at a desk in what appears to be a tech or electronics lab. He is wearing a yellow t-shirt with the word "dil" written on it. He is waving with his right hand....To the right of the laptop, there is another laptop with a screen displaying a graphical interface, possibly for a simulation or a game. There is also a small, colorful robot or electronic device with a screen and buttons on the right side of the desk. The background of the image shows a room with shelves filled with boxes and other items, suggesting that this is a workspace or a lab where various electronic devices and components are stored and used. The floor is tiled, and the overall setting appears to be a tech or electronics workshop or classroom. \\
        \hline
        Object Type & Person \\
        \hline
        Name & Imam \\
        \hline
        User Context & New engineer in the lab \\
        \hline
        Original Scene Image Location & \textit{Path to the scene image location} \\
        \hline
        Cropped Object of Interest Image Location & \textit{Path to the object of interest image location} \\
        \hline
        Retrieval Keywords & 1. Imam, 2. New engineer, 3. Yellow t-shirt, 4. Man waving \newline
        5. Electronics lab, 6. Cluttered desk, 7. Robotics workspace, 8. Smiling engineer \newline
        9. Computer setup, 10. Multiple laptops, 11. Small robots, 12. "dil" shirt \newline
        13. 11:18 time display, 14. Bearded man, 15. Dark-haired engineer, 16. Friendly tech worker \newline
        17. Lab with shelves, 18. Tiled floor workspace, 19. Handheld device, 20. Electronic components \newline
        21. Software interface, 22. Simulation screen, 23. Young man in lab, 24. Approachable engineer \newline
        25. Tech classroom setting \\
        \hline
    \end{tabular}
\end{table*}
\FloatBarrier

\section{Experiments with the HUX AI}
\label{sec:experiments}
\subsection{Goals of the implementation}
Drawing inspiration from the selective attention tests widely used in psychological assessments, which describe how human selective attention performs when there is a task in a dynamic environment. Focusing on a high-cognitive-load task makes it difficult to notice other apparent environmental changes that would otherwise be easily observed.

One notable ability that can be explored using this system is to pay multiple attention to things the user is not paying selective attention to, thus helping out users in real-world stochastic situations. This technology can help verbal recall of objects, temporally and spatially, by having a mere conversation on things that have been overlooked.

\subsection{Test Environment}
The following are knowns and unknowns for the HUX AI.
\textit{Known:} Objects of interest, Events of Interest.
\textit{Unknown:} Sequence of Events, Objects outside of task interest.

The experimental setup is shown in Figure \ref{fig:expt1_view3}. The scene consists of a dark background on a tabletop. The objects of interest are fruits. Additionally, other objects are introduced at different times. The scope of the pipeline restricts its observational capabilities to increase the signal-to-noise ratio by having a limited set of OOIs and EOIs for accomplishing a specific task. For example, for the user to be focused on cooking, the AI should not detect EOIs whenever it identifies random objects it is capable of encountering but rather restrict its event detection capabilities to the foods, food items, and objects related to this specific task. While the VLM is always generalized and can recognize all possible objects, the real-time video analyzer restricts EOI extraction for this specific task. 

With this scope in mind, this particular experiment aims to dynamically change the environment by spatially and temporally changing the OOIs. This includes the removal of fruit from the scene and placing new fruit in the scene in random time sequences. Additionally, none of the language models used were fine-tuned to preserve the system's generality.

\textit{EOIs and OOIs for this experiment:} The OOIs are the fruits in the scene environment. The EOIs are the disappearance and appearance of those fruits (OOIs). 

When the changes in the scene occur, the user also simultaneously interacts with the AI, asking questions about the environment and specific objects in the environment using their eye gaze, questions that need at least one multi-modal data from each of them. This test environment aims to mimic the fundamentals of any environment as a starting point. Promising results in this environment could be applied to other test environments with changes in the OOIs and different task-specific environments. 

We are testing the current implementation with a single user who is informed about the system's capabilities without revealing the technical details. The user was given cues on what questions could be asked to test the system's usefulness and was allowed to improvise. Table \ref{tab:multimodal_questions} represents the different kinds of proposed questions asked to HUX AI while interacting with this environment, which involves multiple modalities that can be utilized for answering different user queries.

\subsection{Hardware Used}

Technical details of the hardware used in the experiment are as follows. We had to divide the computation across three independent systems whose input-output was networked to create the HUX AI system for this test environment.

\subsubsection{System-1 (User Interface):}
This system had the following specifications: 

\begin{itemize}
\item Operating System: Windows 11 Pro
\item RAM: 32 GB
\item vRam: 12 GB
\item Graphics Card: Nvidia RTX 4070 Ti
\item CPU: Intel Core i7 13th gen
\item Eye Tracker: Tobii Eye Tracker 5 
\item Display: 24" with a resolution of 2560 x 1440 pixels
\end{itemize}

\textbf{Assigned responsibilities:}
\begin{itemize}
\item {\texttt{Speech}}: Collects user speech input and converts them to text.
\item {\texttt{User Interface}}: User views the environment through this computer.
\item {\texttt{Eye Gaze}}: Collects user eye gaze from the gaze hardware.
\item {\texttt{Eye Gaze to ROI image}}: Converts raw eye gaze coordinates in (x,y) to ROI images.
\item {\texttt{Scene Capture}}: The camera that captures the scene in real-time is directly connected to this computer.
\item {\texttt{Event Detection}}: The model that is responsible for detecting the event is run in this computer. The events are captured as frames and sent for processing in real time. The hybrid approach, as depicted in Figure \ref{fig:video_filter_3}, was not used in this implementation.
\end{itemize}

\subsubsection{System-2 (Main Hub):} 
This system had the following specifications: 
\begin{itemize}
\item Operating System: Ubuntu 22
\item RAM: 32 GB
\item vRam: 16GB
\item Graphics Card: Nvidia RTX 4070 Ti Super
\item CPU: Intel Core i9-14th Gen
\end{itemize}

\textbf{Assigned responsibilities:}
\begin{itemize}
\item {\texttt{ROI Image}}: ROI gaze image is passed through VLM to have image captions whenever there is a new speech input.
\item {\texttt{Data Collection in real-time}}: All multi-modal data is stored as text here.
\item {\texttt{Data fusion with LLM prompting}}: All the multi-modal data inferred as text is fed into the LLM context window with a prompting strategy. All the data modalities are labeled with their source and their roles. All the data modalities are prompt-engineered to answer the main user query. 
\item {\texttt{Text to Speech}}: Converts the LLM generated text answer to speech output.
\end{itemize}

\textbf{Communication:} Receives Gaze Images and Speech-to-Text converted utterances from System-1 to System-2.

\subsubsection{System-3 (Scene Processing):} 
This system had the following specifications: 
\begin{itemize}
\item Operating System: Ubuntu 22
\item RAM: 32 GB
\item vRam: 12GB
\item Graphic Card: Nvidia RTX 4070 Ti
\item CPU: Intel Core i7 13th gen
\end{itemize}

\textbf{Assigned responsibilities:}
\begin{itemize}
\item {\texttt{Significant Event Scene to Text Description}}: Since the VLM consumes a lot of video memory, the scene processing is outsourced to this computer.
\end{itemize}

\textbf{Communication:} Receives raw EOI frames from System-1, processes them, and sends them to System-3 as image captions in real-time.

\textbf{Deployment limitations:} VLMs and LLM engines are compute-heavy, so has to be deployed in three different systems in parallel and act in synchronization to run the system to the total ability.

\textit{Note:} Multi-modal contextual memory was not integrated in this pipeline rather tested separately (Section \ref{subsec:multimodalcontext}).

\subsection{User Interaction with the AI} 
The user interacts with the AI as follows:

\begin{itemize}
\item  The user presses and holds the space bar in an QWERTY keyboard. The system receives the voice input until the space bar is released.

\item The user can speak anything while holding the space bar, and the voice is captured.

\item  Current gaze image is captured when the space bar is released from hold.

\item  The EOI frames extraction operates independently. They only change when there is a change in the environment, regardless of whether the user presses the space bar.

\item  User waits for the answer after being processed by the pipeline. The user cannot interfere with or pause the HUX AI response for this prototype version.
\end{itemize}

\subsubsection{Noted Errors}
During the trial user interaction runs, there were errors. Contributors of the errors:

\begin{itemize}
\item  Error in the speech-to-text conversion.

\item  User gaze being outside the scope of scene environment: One such occurrence occurred when the user looked at the taskbar of the Windows 11 screen. HUX AI still answered the question, but the response was irrelevant to the scene. 

\item  Recollection from the Context Window: General problems identified in the LLM conversation include hallucination and forgetfulness.
\end{itemize}

\subsection{Multi-modal Context Processing Prompt Architecture}
\begin{itemize}
    \item \textbf{Based on the previous responses and new multi-modal information, answer the next question:}
    \item \textbf{General detailed description of user environment:} \textless Latest Environment Description \textgreater
    \item \textbf{User is specifically gazing at:} \textless Latest Eye-Gaze based ROI Description \textgreater
    \item \textbf{Human's speech query:} \textless Speech to Text Converted Context Rich User Utterance \textgreater
    \item \textbf{Give only a relevant and precise answer to the Human's speech query:} \textless Response \textgreater
\end{itemize}
\subsection{The Experimental Demonstration}
Table \ref{tab:multimodal_questions} indicates the example user queries that can be asked to this implemented system. It also indicates the origin or source of data that the multi-modal context processing can use to answer the queries.

\FloatBarrier
\begin{table*}[h]
\centering
  \caption{Examples of user prompts and the multimodal data used to answer the prompts}
  \label{tab:multimodal_questions}

  \begin{tabular}{p{5cm}cccc}
    \toprule
    User Questions & Gaze & Scene & Speech   & Memory\\
         &  &  &   (Context) & \\
    \midrule
    \\"Do you see any fruits?" & &\checkmark &\
    \\"What are we looking at?" &\checkmark &\
    \\"What are some recipes with it?" &\checkmark &&\checkmark\
    \\"Do you observe anything change in the environment?" &\checkmark &\checkmark\
    \\"Can you repeat the recipe solution you gave me again?" &&&&\checkmark\
    \\"Go through all the multi-modal data you collected during our conversation?" &&&&\checkmark\
    \\"Does this go well with a Pizza?" &\checkmark &&\checkmark\
    \\"My friend John likes this a lot" &\checkmark &&\checkmark\
    \\"Do you think John likes this?" &\checkmark &&&\checkmark\\

  \bottomrule
\end{tabular}
\end{table*}
\FloatBarrier

\begin{figure*}
    \centering
    \includegraphics[width=1\linewidth]{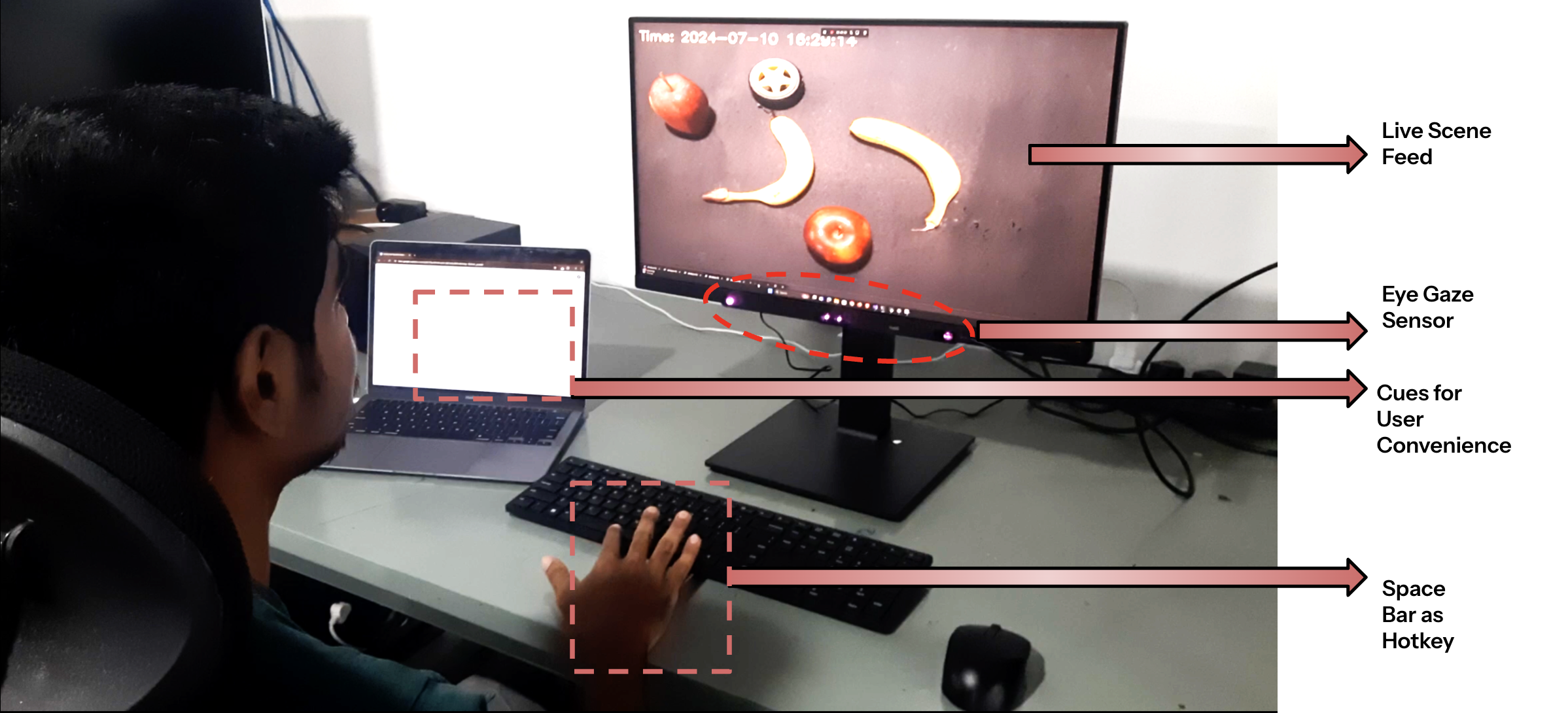}
    \caption{Experiment in the isometric view.}
    \label{fig:expt1_view1}
\end{figure*}

The depicted sources include scene, gaze, speech (new Context), and memory from context windows like previously generated content or previously inferred data. Figure \ref{fig:expt1_view1} shows the experiment setup. This depicts the user seated in front of the display, which shows the real-time scene feed. The text on the laptop is supportive material for the user to test out the capabilities and kinds of queries to be asked (Not to force the test user to remember). Figure \ref{fig:expt1_view1} also shows the usage of the space bar as a hotkey to trigger. Creating new speech input and extracting gaze data at this point will be used to generate ROI. The user utters the queries when holding the space bar pressed, during which the gaze data from the hardware is also extracted.

\begin{figure*}
    \centering
    \includegraphics[width=1\linewidth]{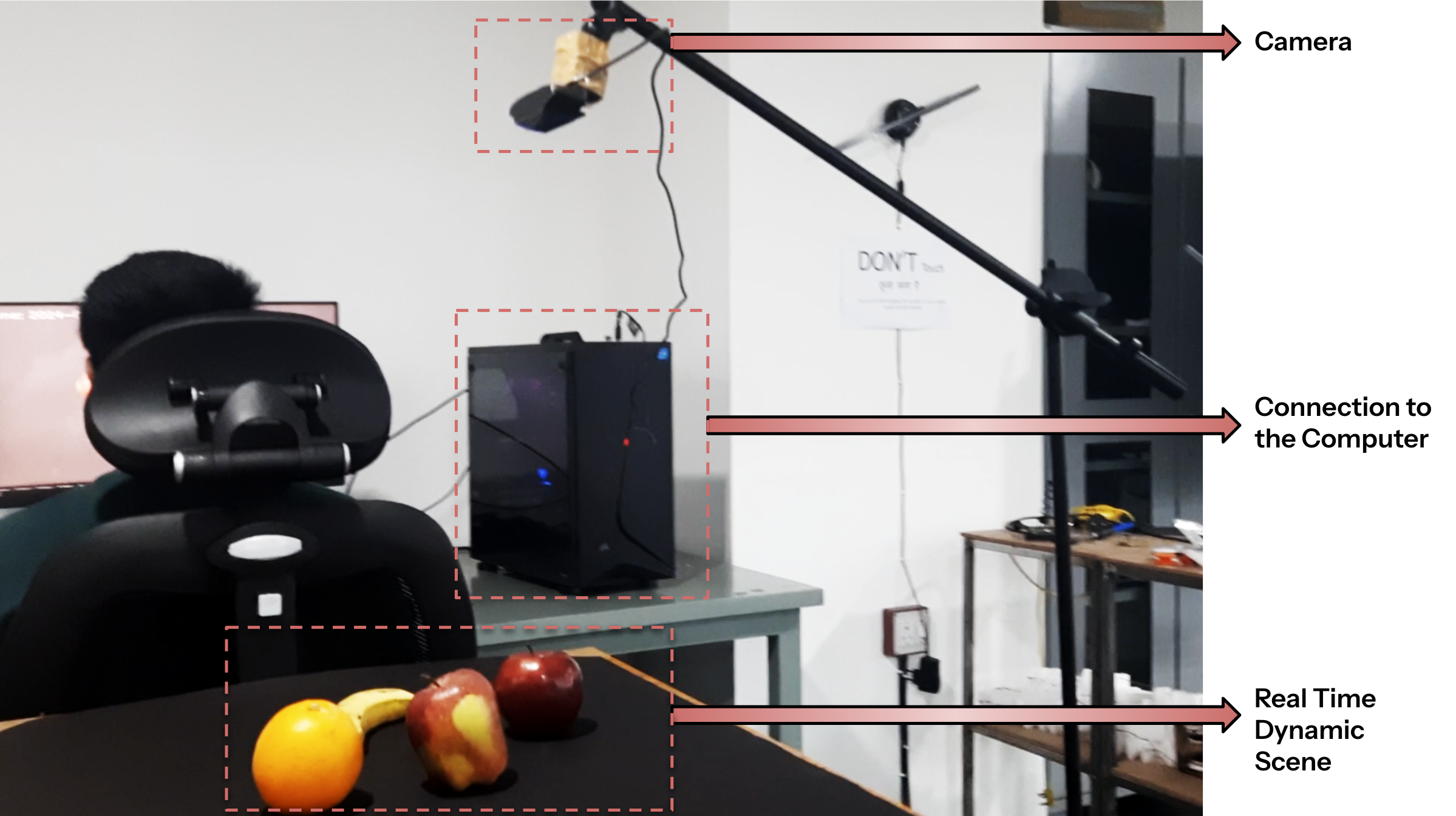}
    \caption{View of the experiment from behind}
    \label{fig:expt1_view2}
\end{figure*}

\begin{figure}
    \centering
    \includegraphics[scale=0.25]{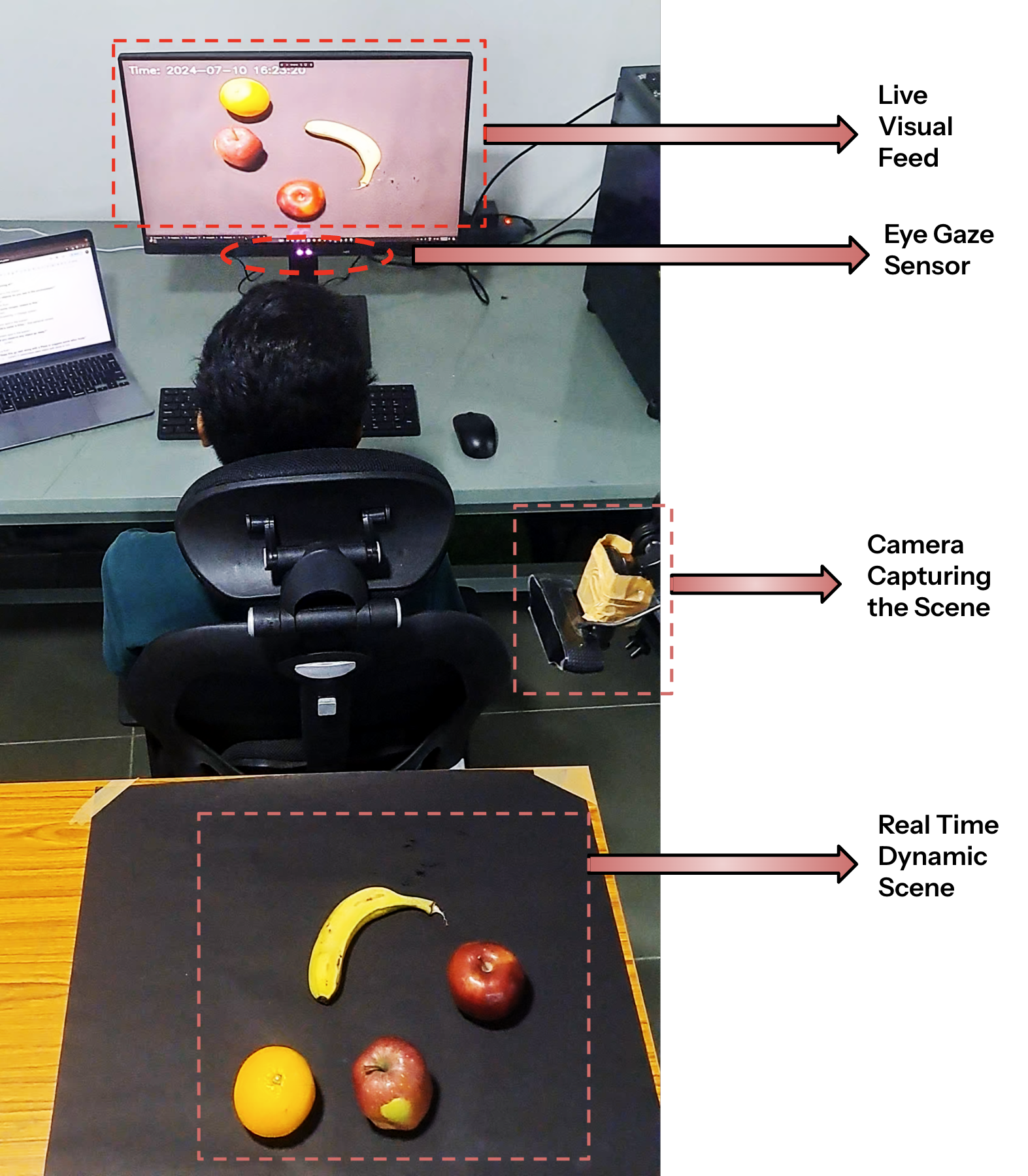}
    \caption{The complete scene where the experiment is conducted}
    \label{fig:expt1_view3}
\end{figure}

Figure \ref{fig:expt1_view2} and \ref{fig:expt1_view3} show the experimental setup in different viewing angles. A webcam is used to capture the live scene. It is connected to the computer as shown in Figure \ref{fig:expt1_view2}. The prompt architecture used in the LLM is responsible for multi-modal Context processing as shown in Table \ref{tab:multimodal_questions}. The consent form was obtained from the user before they were made to work with the HUX AI.
 
\section{Discussions}
\label{sec:discuss}
The issues humans face in selective attention can be overcome using our HUX AI through human-AI collaboration. This is done by retrieving the data from HUX AI memory when humans prompt it in real-time. When used in real-world applications, this feature reduces the cognitive load on the user and helps improve the quality of life of humans. Although our current prototype has latency issues, this can be tackled in future versions of the HUX AI. 

The HUX AI enhances and perceives the scene to aid human perception and ask specific questions and answers regarding the enhanced visual feed. Improving visual information and facilitating interactive questioning helps humans by offering contextual insights, clarifying visual details, and enabling informed real-time decision-making.  

The current method for storing memory is restricted to singular modalities like storing images, texts, or voices in real-time. We demonstrated how HUX AI can create multi-modal context-rich memory and retrieve it through contextual cues. This is useful to humans as it allows for a comprehensive and nuanced understanding of past events by integrating multiple data types. It facilitates accessible and intuitive information retrieval, enhancing memory recall and decision-making based on interconnected data.

\section{Limitations and Future Works}
\label{sec:lim_fut}
\subsection{Current Limitations}
More environments have to be tested to determine where users find them useful. Environments with different objects and complicated events of interest must be tested. Our work is just a starting point for an emerging phenomenon of human computer environment interaction to enhance human perception abilities. They assess multi-modal memory retrieval capabilities in large-scale data comparable to a daily user storage amount. The stack implementation of storage of scene captions leads to loss of scene data (considerable) in real-time interaction.

\subsection{Future Works}

\begin{itemize}
\item {\texttt{Multi-modal Memory Retrieval}}: Creation has been demonstrated in the paper, assessing multi-modal memory retrieval capabilities in large-scale data comparable to a daily user storable amount. The efficacy of the system will be subjected to further evaluation in future also considering privacy limitations.

\item {\texttt{More Test Environments}}: We plan to increase the complexity of the environment by introducing different events and objects of interests.

\item {\texttt{Robotics}}: We plan to conduct further studies on human robot interaction with HUX AI. 

\item {\texttt{Wearable Robotics}}: Wearable robots are on a boom. It can be like extra limbs and human-AI collaboration. Many form factors can be made, such as a snake robot and a manipulator \cite{rajesh2024serpentine}. Figure \ref{fig:HUX_snake_robot} describes how the snake robot is deployed as a wearable robot like a backpack and can help out the user in seamless communication and manipulation tasks when integrated with the HUX AI system.

\begin{figure}
    \centering
    \includegraphics[scale=0.07]{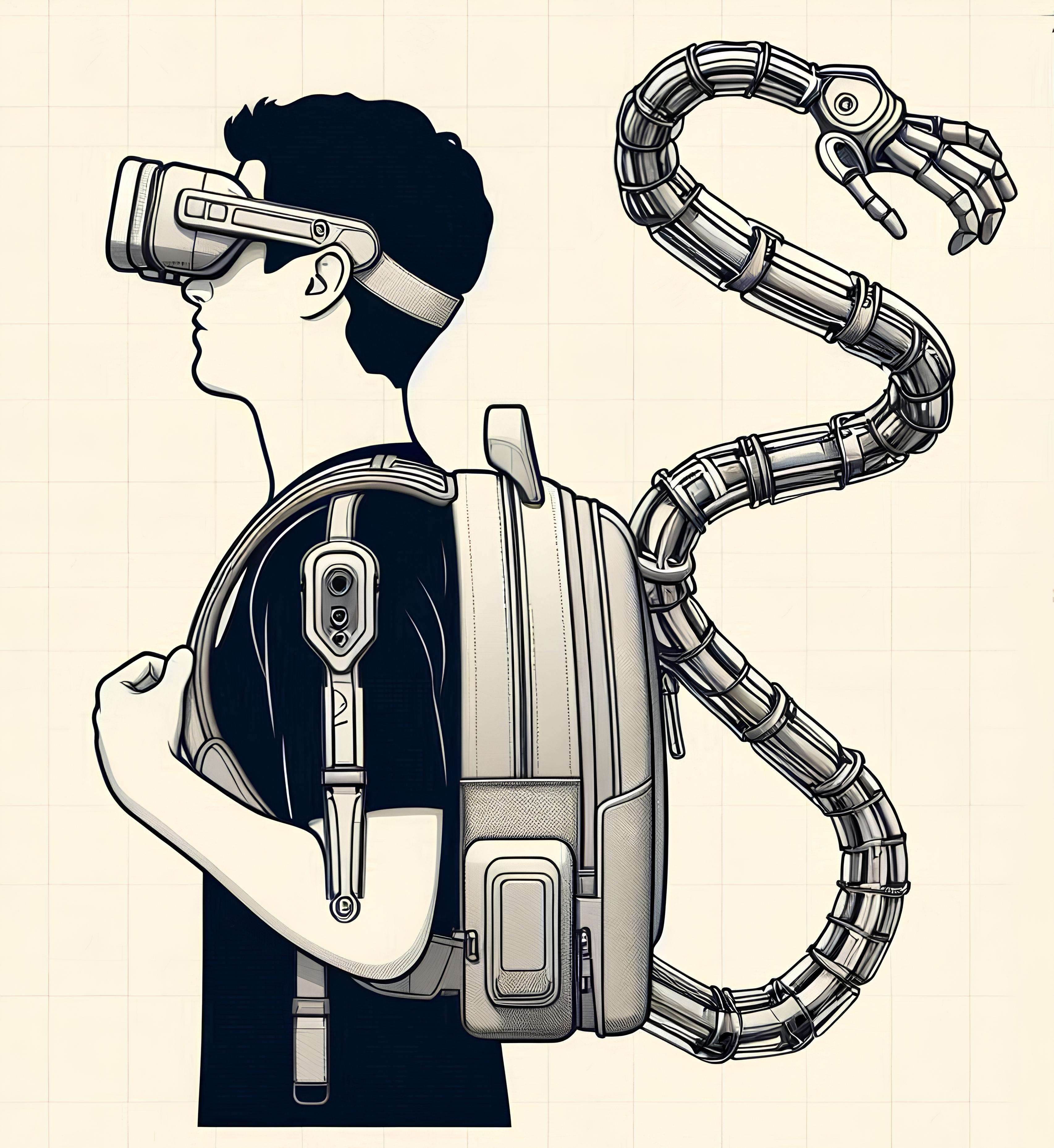}
    \caption{Concept Art of HUX AI coupled with a wearable robot as generated by DALL-E}
    \label{fig:HUX_snake_robot}
\end{figure}

\item {\texttt{Human-AI powered drone collaboration}}: HUX AI that understands human perception uses drones to aid human environment interaction. Additional perceptions and manipulations \cite{vs2021robust} can also be performed.

\item {\texttt{Teleoperated Robotics}}: The camera feed can even be sourced from a teleoperated robot, allowing HUX AI to interact with and perceive the environment through these remotely controlled robots.

\item {\texttt{Smart Appliances}}: Incorporating smart devices and connecting them to the HUX AI system can facilitate seamless control of those devices by understanding user intents. Advanced methods, such as gaze-based actions using spatial reasoning, can be employed to manipulate these devices in familiar and unfamiliar environments. This integration enhances user interaction with smart devices, making it more intuitive and efficient by leveraging the HUX AI system's ability to interpret and act on user intentions.

\item {\texttt{GUI navigation}}: Navigating GUI environments, such as external computer screens, smartphones, smart devices, and VR screens, involves interacting with UI elements and events of interest. Vision-language models (VLM) in GUI environments have been explored in \cite{hong2024cogagent}. Implementing these models in multi-modal systems like HUX AI presents an exciting and valuable problem statement.

\item {\texttt{Empowering Differently Challenged Users}}: Since HUX AI can understand multiple modalities, it can be customized to process multi-modal data according to each user's unique needs. This capability allows it to accurately interpret user intents and facilitate a frictionless Human Computer Environment Interaction, enhancing human-AI collaboration.

\item {\texttt{Aiding Creativity and Brainstorming}}: The ability to store context-rich, multi-modal memories enables the fusion of diverse data types, greatly aiding creative endeavors such as brainstorming. This approach mirrors human creativity, combining different insights to form singular, innovative ideas. HUX AI can support more holistic and imaginative problem-solving processes by integrating and synthesizing varied information.
\end{itemize}

\section{Conclusion}
\label{sec:conclu}
In this work, we introduced the Heads-Up eXperience (HUX) artificial intelligence system, which aims to revolutionize the landscape of personal smart devices. HUX AI is designed as a personal AI to enhance human computer environment interaction (HCEI) by leveraging its ability to detect changes in task-specific environments and remember multi-modal, context-rich short and long-term memories. This design promotes seamless human-AI collaboration across various scenarios, fundamentally altering how humans interact with computers and their environment. HUX AI enables a sophisticated multi-modal interaction framework by incorporating eye gaze, scene analysis, and speech input. This approach facilitates more effective and intuitive interactions and sets the stage for future advancements in personalized intelligent technologies.

\section*{References}

\bibliography{mybibfile}

\end{document}